\documentclass[a4paper, 11pt]{article} 
\usepackage[english]{babel}
\usepackage{amsfonts}
\usepackage{amsmath}
\usepackage{amssymb}
\usepackage{array}
\usepackage{dcolumn}
\usepackage[mathcal]{eucal}
\usepackage[utf8]{inputenc}
\usepackage{geometry}
\usepackage{multicol}
\usepackage{psfrag}
\usepackage[dvips]{graphicx}
\usepackage{vmargin} 
\usepackage[ec]{aeguill}
\usepackage{enumerate} 
\usepackage{version} 
\usepackage[T1]{fontenc}
\usepackage{url}
\usepackage{indentfirst}
\usepackage{setspace}
\usepackage{fancybox}
\usepackage{cases}
\usepackage{xcolor} 
\usepackage[pdfborder={0 0 0}]{hyperref}
\usepackage{ulem}

\usepackage{sectsty}

\usepackage{tikz}
\usepackage{longtable}

   \newenvironment{changemargin}[2]{\begin{list}{}{%
   \setlength{\topsep}{0pt}%
   \setlength{\leftmargin}{0pt}%
   \setlength{\rightmargin}{0pt}%
   \setlength{\listparindent}{\parindent}%
   \setlength{\itemindent}{\parindent}%
   \setlength{\parsep}{0pt plus 1pt}%
   \addtolength{\leftmargin}{#1}%
   \addtolength{\rightmargin}{#2}%
   }\item }{\end{list}}

\setmarginsrb{3cm}{2cm}{3cm}{2cm}{0cm}{1cm}{0cm}{1cm} 

\title{\bf Testing quantum-like models of judgment for question order effect}

\author{Thomas Boyer-Kassem\footnote{
Archives H. Poincaré (UMR 7117: CNRS, Université de Lorraine), 91 avenue de la Libération, 54000 Nancy, France, and Tilburg Center for Logic, General Ethics and Philosophy of Science, Tilburg University, PO Box 90153, 5000 LE Tilburg, The Netherlands. E-mail: \href{mailto:t.c.e.boyer-kassem@uvt.nl}{t.c.e.boyer-kassem@uvt.nl}}, Sébastien Duchêne\footnote{GREDEG (UMR 7321: CNRS, Université de Nice Sophia Antipolis), 250 rue Albert Einstein, 06560 Valbonne, France. E-mail: \href{mailto:sebastien.duchene@gredeg.cnrs.fr}{sebastien.duchene@gredeg.cnrs.fr}, \href{mailto:eric.guerci@gredeg.cnrs.fr}{eric.guerci@gredeg.cnrs.fr}}, Eric Guerci\textsuperscript{$\dagger$}\\
\vspace{5mm}\\
to appear in \textit{Mathematical Social Sciences}, 80 (2016): 33-46.\\
\href{http://dx.doi.org/10.1016/j.mathsocsci.2016.01.001}{http://dx.doi.org/10.1016/j.mathsocsci.2016.01.001}
}

\date{}

\begin{document}

\maketitle

\begin{abstract}
Lately, so-called ``quantum" models, based on parts of the mathematics of quantum mechanics, have been developed in decision theory and cognitive sciences to account for seemingly irrational or paradoxical human judgments. We consider here some such quantum-like models that address question order effects, i.e. cases in which given answers depend on the order of presentation of the questions.
Models of various dimensionalities could be used; can the simplest ones be empirically adequate?
From the quantum law of reciprocity, we derive new empirical predictions that we call the Grand Reciprocity equations, that must be satisfied by several existing quantum-like models, in their non-degenerate versions. Using substantial existing data sets, we show that these non-degenerate versions fail the GR test in most cases, which means that, if quantum-like models of the kind considered here are to work, it can only be in their degenerate versions. However, we suggest that the route of degenerate models is not necessarily an easy one, and we argue for more research on the empirical adequacy of degenerate quantum-like models in general.
\end{abstract}


\bigskip

\begin{flushright}
\textit{``Research is needed which designs tests of quantum properties\\ such as the law of reciprocity and the law of double stochasticity" \\
(Busemeyer and Bruza 2012, 342)}
\end{flushright}

\section{Introduction}

In decision theory and in cognitive sciences, classical cognitive models of judgment rely on Bayesian probabilities and suppose that agents' decisions or choices are guided by preferences or attitudes 
that are determined at any time. Yet, various empirical results have threatened the predictive and explanatory power of these classical models: human judgments display order effects --- the answers given to two questions depend on the order of presentation of these questions (Schuman and Presser 1981, Tourangeau, Ribs and Rasinski 2000, Moore 2002) ---, conjunction fallacies --- someone is judged less likely to have characteristic $C$ than characteristics $C$ and $D$ (Tversky and Kahneman 1982 and 1983, Gavanski and Roskos-Ewoldsen 1991, Stolarz-Fantino \textit{et al.} 2003) ---, violate the sure-thing principle --- stating that preferring $X$ to $Y$ given any possible state of the world should imply preferring $X$ to $Y$ when the exact state of the world is not known 
(Allais 1953, Ellsberg 1961, Shafir and Tversky 1992) ---, or asymmetries in similarity --- $X$ is judged more similar to $Y$ than $Y$ to $X$ (Tversky 1977, Krumhansl 1978, Ashby and Perrin 1988).
In the classical cognitive framework, these behaviors have often been dubbed as irrational or paradoxical.

Recently, a research field that rely on so-called ``quantum" or ``quantum-like" models has developed to account for such behaviors.
The qualifier ``quantum" is used to indicate that the models exploit the mathematics of a contemporary physical theory, quantum mechanics. 
Note that only some mathematical tools of quantum mechanics are employed, and that the claim is not 
that these models are justified by an application of quantum physics to the brain. 
For that reason, we shall prefer to call them ``quantum-like" models.
Such models put into question two classical characteristics recalled above: they abandon Bayesian probabilities for others which are similar to probabilities in quantum mechanics, and they allow for preferences or attitudes to be undetermined. 
Quantum-like models have received much interest from psychologists, physicists, economists, cognitive scientists and philosophers. For example, new theoretical frameworks have been proposed in decision theory and bounded rationality (Danilov and Lambert-Mogiliansky 2008 and 2010, Yukalov and Sornette 2011).

Various quantum-like models have been proposed to account for any of the seemingly paradoxical behaviors recalled above (for reviews, see Pothos and Busemeyer 2013, Ashtiani and Azgomi 2015). 
First, as the mathematics involved in quantum mechanics are well-known for their non-commutative features, one of their natural application is the account of question order effects. For instance, Wang and Busemeyer (2009) and Wang \textit{et al.} (2014) offer a general quantum-like model for attitude questions that are asked in polls. 
Conte \textit{et al.} (2009) present a model for mental states of visual perception, and Atmanspacher and Römer (2012) discuss non-commutativity. 
Quantum-like models have also been offered to explain the conjunction fallacy. For instance, Franco (2009) argues that it can be recovered from interference effects, which are central features in quantum mechanics and at the origin of the violation of classical probabilities. 
Busemeyer \textit{et al.} (2011) present a quantum-like model that could explain conjunction fallacy from some order effects. 
The violation of the sure thing principle has also been investigated by means of quantum-like models. Busemeyer, Wang and Townsend (2006), and Busemeyer and Wang (2007) 
use quantum formalism to explain this violation by introducing probabilistic interference and superposition of states. Khrennikov and Haven (2009) also explain Ellsberg's paradox, and Aerts \textit{et al.} (2011) model the Hawaii problem. 
Lambert-Mogiliansky \textit{et al.} (2009) offer a model for the indeterminacy of preferences.
Dynamical models, that rely on a time evolution of the mental state, have also been introduced (Pothos and Busemeyer 2009, Busemeyer \textit{et al.} 2009, and Trueblood and Busemeyer 2011).
Several other empirical features, such as asymmetry judgments in similarity 
have also been offered a quantum-like model (Pothos and Busemeyer, 2011).

We shall concentrate in this paper on one kind of quantum-like models, those which claim to account for question order effect --- that answers depend on the order in which questions are asked. We choose these models because (i) order effect is a straightforward application of the non-commutative mathematics from quantum mechanics, (ii) order effect models are perhaps the simplest ones and (iii) some of them are at the roots of other models.
All quantum-like models for order effect are not identical. Yet, several of them are built along the same lines, as in Conte \textit{et al.} (2009), Busemeyer \textit{et al.} (2009), Busemeyer \textit{et al.} (2011), Atmanspacher and Römer (2012), Pothos and Busemeyer (2013), Wang and Busemeyer (2013) and Wang \textit{et al.} (2014). 
We choose to focus here on this kind of models for order effect (to be characterized, at least partly, in Section~2). So, we do not consider for instance the model in Aerts, Gabora and Sozzo (2013) that rely on different hypotheses, or a quantum-like model for order effect in inference that can be found in Trueblood and Busemeyer (2011) or in Busemeyer \textit{et al.} (2011). Our analyses only bear on the previously mentioned kind of models.

When such quantum-like models for order effects were proposed, empirical adequacy was cared about:
it was argued that the models were able to account either for existing empirical data (e.g. Wang and Busemeyer 2013 for the data from Moore 2002, Wang \textit{et al.} 2014 for data from around 70 national surveys) or for data from new experiments (e.g. Conte \textit{et al.} 2009, Wang and Busemeyer 2013). This is not all: a supplementary \textit{a priori} constraint (the ``QQ equality")  has been derived 
and it has been successfully verified on the above-mentioned data (Wang and Busemeyer 2013, Wang \textit{et al.} 2014). 
So, it seems that quantum-like models for question order effects are well verified and can be considered as adequate, at least for a vast set of experiments.

However, some details remain to be settled. 
In many cases, the above-referred papers do not specify exactly which model, with which parameters, is to account for the data in this or thus order effect situation. 
In particular, when it is claimed that some data can be accounted for by a quantum-like model, the dimensionality of the model is often not specified, and the most general case is assumed: the model is supposed to be $N$ dimensional, and the answers to the questions can be represented by subspaces of any dimension.
This encompasses two cases: either the answers are represented by 1-D subspaces, in which case the model is said to be \textit{non-degenerate}, or the answers are represented by subspaces of larger dimensions, in which case the model is said to be \textit{degenerate}.
The papers do not always specify whether the model that is to account for the data can remain simple and be non-degenerate, or has to be degenerate --- this is a question of theoretical importance, as we shall argue.
The quantum-like literature is inclined to favor 
degenerate models, insisting on the high-dimensionality of the space and on the complexity of the experimental situation to be modeled, and often considers non-degenerate models as toy models and use them for illustrative purposes only%
\footnote{\label{footnote_toy-model}For instance, in Busemeyer and Bruza (2012), the non-degenerate model is presented as ``a simple example" (p. 100) or as ``much too simple" (p. 108), before the general model, which is degenerate, is presented. That non-degenerate models are ``toy models" is also what suggest Busemeyer \textit{et al.} 2015 p. 240, albeit for quantum-like models for the conjunction fallacy.
An exception towards degeneracy is Conte \textit{et al.} (2009), one of the first papers in the field, which only considers non-degenerate models.}.
So, the question of choosing between non-degenerate and degenerate models on an empirical basis has not been fully dealt with, and it remains unclear so far whether the simple non-degenerate version of the models could be empirically adequate.
Is there a simple way to test whether a non-degenerate version of the quantum-like models is sufficient to account for some order effect data?

The present paper argues that this question can receive a positive answer. We show that the non-degenerate versions of the quantum-like models under consideration can actually be empirically tested in another unstudied --- although simple --- way, that bears on data that were not intended to be explained but that are nevertheless predicted by the models. After reconstructing a non-degenerate quantum-like version of a model for order effect (Section~2), we derive from the quantum law of reciprocity a set of constraints for it that we call the Grand Reciprocity (GR) equations (Section~3). 
In Section~4, we argue that the non-degenerate versions of the models that can be found in Conte \textit{et al.} (2009), Busemeyer and Bruza (2012), Pothos and Busemeyer (2013), Wang and Busemeyer (2013) and Wang \textit{et al.} (2014), have to obey the GR equations --- note that these papers, except for Conte \textit{et al.} (2009), also consider degenerate versions of the models, that are not required to obey the GR equations, and thus will not be tested here.
Using only available data and without carrying out any new experiment, we are able to put the GR equations to the test (Section~5).  We show that a vast majority of empirical cases fail to satisfy the GR equations, which means that the non-degenerate versions of the above-mentioned models cannot account for these cases. In other words, if the quantum-like models that have been proposed are indeed to work, it cannot be in the special case of their non-degenerate versions, as the authors themselves had rightly anticipated, and the data should be accounted for with degenerate models.
In Section~6, we suggest however that the route of degenerate models is not necessarily an easy one, and we argue for more research on the empirical adequacy of models in general.
Overall, the positive conclusion of our results, we suggest, is that research on quantum-like models for question order effect should be directed towards degenerate models, both as a consequence of the failure of non-degenerate ones and in order to find generalizations of the GR equations that should enable to empirically test them.
This paper is the first one to challenge non-degenerate models for order effect and to show that they cannot be used as such in many cases. Our work puts warnings about the future use of these specific models, and provides a critical and constructive look at quantum models of cognition.

\section{A general non-degenerate quantum-like model for question order effect}
\label{sec_general-quantum-model}

As indicated in the introduction, we are concerned here with the following quantum-like models for order effect: Conte \textit{et al.} (2009), Busemeyer \textit{et al.} (2009), Busemeyer \textit{et al.} (2011), Atmanspacher and Römer (2012), Pothos and Busemeyer (2013), Wang and Busemeyer (2013) and Wang \textit{et al.} (2014).
As they are more or less built along the same lines, 
we choose for simplicity to present in this section one simple and general model, which sets the notations and on which all the discussions will be made. 
A crucial point is this: these models, except for Conte \textit{et al.} (2009), are quite general and can be degenerate or non-degenerate, as already noted. As we shall be concerned here with testing their non-degenerate versions only, as a special case, the model we present in this section is non-degenerate. 
In Section~4, we shall discuss in detail how this model is included in the models from the above list.

The model is about the beliefs of a person, about which dichotomous yes-no questions can be asked, for instance ``is Clinton honest?". 
A vector space on the complex numbers is introduced to represent the beliefs of the agent, and the answers to the questions.
In the model, two dichotomous questions $A$ and $B$ are posed successively to an agent, in the order $A$-then-$B$ or $B$-then-$A$. Answer ``yes" (respectively ``no") to $A$ is represented by the vector $\vert a_0 \rangle$ (respectively $\vert a_1 \rangle$), and similarly for $B$, with vectors $\vert b_0 \rangle$ (respectively) $\vert b_1 \rangle$.
It is important to note that an answer is supposed to be represented by a vector (or more exactly, see below, by the ray defined by this vector), and not by a plane or by any subspace of dimension greater than 1.
Since there are no other possible answers to question $A$ (respectively $B$) 
than 0 and 1, the set ($\vert a_0 \rangle$, $\vert a_1 \rangle$) (respectively ($\vert b_0 \rangle$, $\vert b_1 \rangle$)) 
forms a basis of the vector space of possible answers, and the vector space is thus of dimension $2$. Note that it is the same vector space that is used to represents answers to questions $A$ and $B$; this space just has two different bases.\footnote{In the literature, questions that can be represented in this way are known as ``incompatible". ``Compatible" questions have a common basis, but the model is then equivalent to a classical one, and there is nothing quantum about it. Cf. for instance Busemeyer and Bruza (2012, 32--34).}

\begin{figure}[tb]
\begin{center}
\setlength{\unitlength}{.7mm}
\begin{picture}(100,80)(-30,-10) 
\put(0, 0){\vector(1,0){50}}
\put(52, -2){$\vert a_0 \rangle$}
\put(0, 0){\vector(0,1){50}}
\put(-5, 55){$\vert a_1 \rangle$}
\put(0, 0){\vector(3,1){47.4}} 
\put(50, 15){$\vert b_0 \rangle$}
\put(0, 0){\vector(-1,3){15.8}} 
\put(-25, 50){$\vert b_1 \rangle$}
\qbezier(30,0)(30,5)(28.4,9.4)
\put(28.4, 9.4){\vector(-1,4){.01}}
\put(32,3){$\delta$}
\qbezier(0,30)(-5,30)(-9.4,28.4)
\put(-9.4, 28.4){\vector(-4,-1){.01}}
\put(-7,32){$\delta$}
\qbezier[20](47.4,0)(47.4,7.6)(47.4, 15.8)
\put(35,-6){$\langle a_0 \vert b_0 \rangle$}
\qbezier[20](-15.8,47.4)(-7.6,47.4)(0, 47.4)
\put(1,45){$\langle a_1 \vert b_1 \rangle$}
\qbezier[60](47.4,15.8)(23.7,15.8)(0, 15.8)
\put(1,18){$\langle a_1 \vert b_0 \rangle$}
\qbezier[60](-15.8,47.4)(-15.8,23.7)(-15.8,0)
\qbezier[20](-15.8,0)(-7.9,0)(0,0)
\put(-25,-6){$\langle a_0 \vert b_1 \rangle$}
\end{picture}
\qquad
\begin{picture}(100,80)(-30,-10) 
\put(0, 0){\vector(1,0){50}}
\put(52, -2){$\vert a_0 \rangle$}
\put(0, 0){\vector(0,1){50}}
\put(-5, 55){$\vert a_1 \rangle$}
\put(0, 0){\vector(3,1){47.4}} 
\put(50, 15){$\vert b_0 \rangle$}
\put(0, 0){\vector(-1,3){15.8}} 
\put(-25, 50){$\vert b_1 \rangle$}
\thicklines 
\put(0, 0){\vector(4,3){40}} 
\put(42, 33){$\vert \psi \rangle$}
\thinlines
\qbezier[30](40,0)(40,15)(40,30)
\put(38, -7){$\alpha_0$}
\qbezier[40](0,30)(20,30)(40,30)
\put(-7, 30){$\alpha_1$}
\qbezier[10](45,15)(42.5,22.5)(40,30)
\put(44, 9){$\beta_0$}
\qbezier[30](-5,15)(17.5,22.5)(40,30)
\put(-12, 10){$\beta_1$}
\end{picture}
\end{center}
\caption{[Left:] The basis vectors $\vert b_0 \rangle$ and $\vert b_1 \rangle$ can be decomposed on the other basis vectors $\vert a_0 \rangle$ and $\vert a_1 \rangle$, so as to be expressed as in eq.~\ref{eq_dec_b0} and \ref{eq_dec_b1}. The scalar products are either equal to $\cos \delta$ or to $\sin \delta$. 
[Right:] The state vector $\vert \psi \rangle$ can be expressed in the two orthonormal bases ($\vert a_0 \rangle$, $\vert a_1 \rangle$) and ($\vert b_0 \rangle$, $\vert b_1 \rangle$). These figures assume the special case of a Hilbert space on real numbers.}\label{fig_bases-A-and-B}
\end{figure}
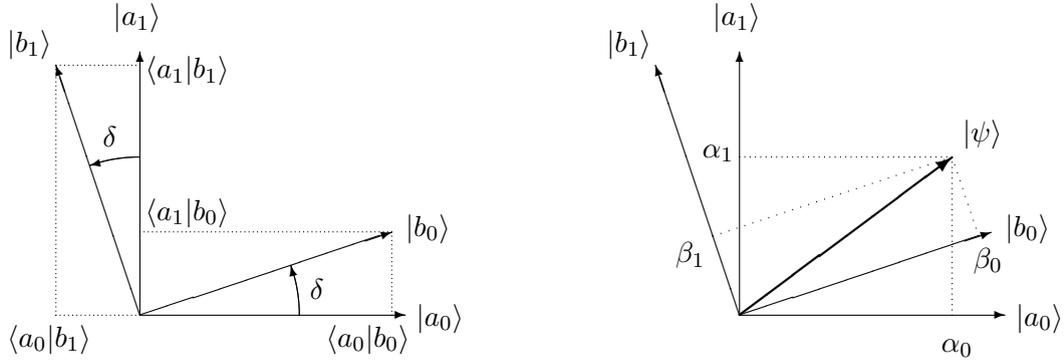

The vector space is supposed to be equipped with a scalar product, thus becoming a Hilbert space: for two vectors $\vert x \rangle$ and $\vert y \rangle$, the scalar product $\langle x \vert y \rangle$ is a complex number; its complex conjugate, $\langle x \vert y \rangle^*$, is just $\langle y \vert x \rangle$. The Hilbert space is on the complex numbers, and vectors can be multiplied by any complex number.\footnote{We consider here this general case. In the literature, some quantum-like models consider a real Hilbert space, in which scalar products are real numbers, and vectors can be multiplied by real numbers only.}
The basis ($\vert a_0 \rangle$, $\vert a_1 \rangle$) is supposed to be orthonormal, i.e. $\langle a_0 \vert a_1 \rangle = 0$ and $\langle a_0 \vert a_0 \rangle = \langle a_1 \vert a_1 \rangle = 1$; similarly for $B$.
Note that there exists a correspondence between the two bases (cf. figure~\ref{fig_bases-A-and-B} left):
\begin{equation} \label{eq_dec_b0}
\vert b_0 \rangle = \langle a_0 \vert b_0 \rangle \vert a_0 \rangle + \langle a_1 \vert b_0 \rangle \vert a_1 \rangle,
\end{equation}
\begin{equation}\label{eq_dec_b1}
\vert b_1 \rangle = \langle a_0 \vert b_1 \rangle \vert a_0 \rangle + \langle a_1 \vert b_1 \rangle \vert a_1 \rangle,
\end{equation}
and similarly for $\vert a_0 \rangle$ and $\vert a_1 \rangle$ expressed as a function of $\vert b_0 \rangle$ and $\vert b_1 \rangle$.

An agent's beliefs about the subject matter of the questions $A$ and $B$ are represented by a normalized belief state
 $\vert \psi \rangle$ from this vector space ($\vert \langle \psi \vert \psi \rangle \vert^2 = 1$). It is supposed to gather all the relevant information to predict her behavior in the situation (like in orthodox quantum mechanics).
$\vert \psi \rangle$ can be expressed in the basis ($\vert a_0 \rangle$, $\vert a_1 \rangle$) as 
\begin{equation}
\vert \psi \rangle = \alpha_0 \vert a_0 \rangle + \alpha_1 \vert a_1 \rangle,
\end{equation}
with ($\alpha_0, \alpha_1) \in \mathbb{C}^2$ (cf. fig.~\ref{fig_bases-A-and-B} right).
It can also be expressed in the basis ($\vert b_0 \rangle$, $\vert b_1 \rangle$) as 
\begin{equation}
\vert \psi \rangle = \beta_0 \vert b_0 \rangle + \beta_1 \vert b_1 \rangle,
\end{equation}
with ($\beta_0, \beta_1) \in \mathbb{C}^2$, and with an appropriate correspondence between the coefficients.

The belief state $\vert \psi \rangle$ determines the answer in a probabilistic way, and changes only when a question is answered, according to the following rules:
\begin{itemize}
\item Born's rule: the probability for the agent to answer $x_i$ ($i= 0,1$) to question $X$ ($X = A, B$) is given by the squared modulus of the scalar product between $\vert \psi \rangle$ and $\vert x_i \rangle$: 
\begin{equation}
\text{Pr}(x_i) =  \vert \langle x_i \vert \psi \rangle \vert^2
\end{equation}
\item projection postulate: the agent's belief state just after the answer $x_i$ is the normalized projection of her belief state prior to the question onto the vector $\vert x_i \rangle$ corresponding to her answer: 
\begin{equation}
\vert \psi \rangle \quad \longmapsto 
\quad \frac{\langle  x_i \vert \psi \rangle}{\vert \langle x_i \vert \psi \rangle \vert}  \vert x_i \rangle.
\end{equation}
\end{itemize}

For instance, if an agent is described by the state vector $\vert \psi \rangle = \alpha_0 \vert a_0 \rangle + \alpha_1 \vert a_1 \rangle$, the probability that she answers $i$ to question $A$ is given by $\vert \alpha_i \vert^2$, in which case the state after the answer is $\frac{\alpha_i}{\vert \alpha_i \vert}  \vert a_i \rangle$. In fig.~\ref{fig_bases-A-and-B}, this probability can be obtained by first orthogonally projecting $\vert \psi \rangle$ on the basis vector corresponding to the answer, and then taking the square of this length.
A consequence of the projection postulate in this model is that, just after an answer $i$ to question $X$ has been given, the state is of the form $\lambda \vert x_i \rangle$ with $\lambda \in \mathbb{C}$ and $\vert \lambda \vert =1$. 
The fact that the state after the answer is equal to $\vert x_i \rangle$ ``up to a phase factor", as one says, is true whatever the state prior to the question. 
In fig.~\ref{fig_bases-A-and-B}, in the case of a real Hilbert space, the projection postulates can be interpreted as follow: project $\vert \psi \rangle$ on the basis vector corresponding to the answer, then normalize it (i.e. expand it so that it gets a length~1); the result is  $\pm \vert x_i \rangle$, according to the relative orientation of $\vert \psi \rangle$  and $\vert x_i \rangle$.
In the general case, answering a question modifies the agent's state of belief, but there are exceptions: if an agent's state of belief is $\lambda \vert x_i \rangle$ (with $\vert \lambda \vert =1$), Born's rule states that she will answer $i$ (``yes" or ``no") with probability~1, and her state of belief is thus unchanged. Such vectors from which an answer can be given with certainty are called ``eigenvectors", and their set is the ``eigenspace", for the ``eigenvalue" $i$. 
In this model, all eigenspaces are of dimension~1, and are called ``rays" (equivalently, one can say that the eigenvalue is not degenerate), since it has been supposed that an answer is represented by a vector, and not by several independent vectors; this will be of decisive importance in the next section.
Another consequence of the projection postulate is that, once an agent has answered $i$ to question $A$, she will answer $i$ with probability~1 to the same question $A$ if it is posed again just afterwards.\footnote{Of course, this is not true if another question, say $B$, is posed in-between.}

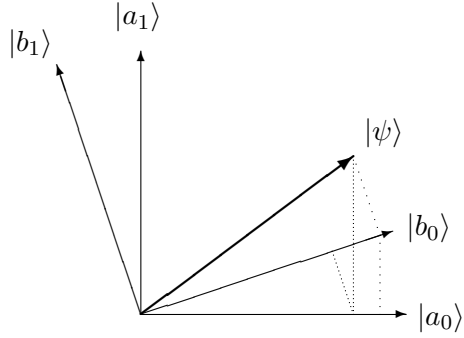
\begin{figure}[!ht]
\begin{center}
\setlength{\unitlength}{.7mm}
\begin{picture}(80,70)(-30,-10) 
\put(0, 0){\vector(1,0){50}}
\put(52, -2){$\vert a_0 \rangle$}
\put(0, 0){\vector(0,1){50}}
\put(-5, 55){$\vert a_1 \rangle$}
\put(0, 0){\vector(3,1){47.4}} 
\put(50, 15){$\vert b_0 \rangle$}
\put(0, 0){\vector(-1,3){15.8}} 
\put(-25, 50){$\vert b_1 \rangle$}
\thicklines 
\put(0, 0){\vector(4,3){40}} 
\put(42, 33){$\vert \psi \rangle$}
\thinlines
\qbezier[40](40,0)(40,15)(40,30)
\qbezier[16](40,0)(38,6)(36,12)
\qbezier[10](45,15)(42.5,22.5)(40,30)
\qbezier[10](45,15)(45,7.5)(45,0)
\end{picture}
\end{center}
\caption{The state vector $\vert \psi \rangle$, projected first on $\vert a_0 \rangle$ and then on $\vert b_0 \rangle$, or first on $\vert b_0 \rangle$ and then on $\vert a_0 \rangle$, gives different lengths. Consequently, the corresponding probabilities $p(a_0,b_0)$ and $p(b_0,a_0)$ are different.}\label{fig_proj-psi-selon-ordre}
\end{figure}

Such a quantum-like model displays order features. Compare for instance $p(a_0,b_0)$, the probability to answer 0 to question $A$ and then 0 to question $B$, and $p(b_0,a_0)$, the probability to give the same answers but in the reverse order. To compute $p(a_0,b_0)$, one can project the initial state on $\vert a_0 \rangle$ without normalizing the result, then project the result on $\vert b_0 \rangle$, still without normalizing, and take the squared modulus of the final result.%
\footnote{Proof. Note 
$p(y_j|x_i)$ 
the probability to answer $j$ to question $Y$ given that question $X$ has been answered with $i$. After the answer $x_i$, the state is $\lambda \vert x_i \rangle$ with $\vert \lambda \vert =1$, so $p(y_j|x_i) = \vert \langle y_j \vert x_i \rangle \vert^2$. Note $p(x_i)$ the probability to answer $i$ to question $X$ when this question is asked first. $p(a_0, b_0) = p(a_0) \cdot p(b_0|a_0) = \vert \langle a_0 \vert \psi \rangle \vert^2 \cdot \vert \langle b_0 \vert a_0 \rangle \vert^2 =  \vert \langle a_0 \vert \psi \rangle \cdot \langle b_0 \vert a_0 \rangle \vert^2 = \vert \langle b_0 \vert a'_0 \rangle \vert^2$, where $\vert a'_0 \rangle =  \langle a_0 \vert \psi \rangle \vert a_0 \rangle$. Since $\vert a'_0 \rangle$ is $\vert \psi \rangle$ projected onto $\vert a_0 \rangle$ and not normalized, computing $\vert \langle b_0 \vert a'_0 \rangle \vert^2$ means that $\vert a'_0 \rangle$ is projected onto $\vert b_0 \rangle$ and not normalized, before the squared modulus is computed. QED.}
In other words, to compare the two probabilities, one can just compare the length of successive projections of $\vert \psi \rangle$, first on $\vert a_0 \rangle$ and then $\vert b_0 \rangle$, or in the reverse order. Figure~\ref{fig_proj-psi-selon-ordre} shows that they are not necessary equal.
Because quantum-like models display these order features, it has been naturally suggested that they can account for experimentally documented order effects; Section~\ref{sec_test-order-effects} discusses several such models.

\section{Constraints: the Grand Reciprocity equations}
\label{sec_reciprocity-relations}

Some general empirical predictions can be derived from the non-degenerate model presented in the previous section.

\subsection{Derivation of the Grand Reciprocity equations}
\label{subsec_derivation-recipro-relations}

Be $x_i$ and $y_j$ the answers an agent gives to two successive questions $X$ and $Y$, in one order or in the other ($X$ may be equal to $Y$, and $i$ to $j$).
Because of the projection postulate, the state just after the answer $x_i$ (respectively $y_j$) is $\lambda \vert x_i \rangle$ (respectively $\lambda' \vert y_j \rangle$), with $(\lambda, \lambda') \in \mathbb{C}^2$ and $\vert \lambda \vert = \vert \lambda' \vert = 1$. 
On the one hand,
\begin{equation}
p(y_j|x_i) = \vert \langle y_j  | (\lambda \vert x_i \rangle) \vert^2 = \vert \langle y_j \vert x_i \rangle \vert^2.
\end{equation}
On the other hand, 
\begin{equation}
p(x_i|y_j) = \vert \langle x_i | (\lambda' \vert y_j \rangle) \vert^2 = \vert \langle x_i \vert y_j \rangle \vert^2.
\end{equation}
As indicated previously, $\langle y_j \vert x_i \rangle = \langle x_i \vert y_j \rangle ^*$, hence $\vert \langle y_j \vert x_i \rangle \vert^2 = \vert \langle x_i \vert y_j \rangle \vert^2$, and so
\begin{equation}\label{eq_law-reciprocity-general}
p(y_j|x_i) = p(x_i|y_j).
\end{equation}

This equation is well-known in quantum mechanics, and is called the law of reciprocity (cf. Peres 1993, p. 35--36 and 56).
The only condition for this law is that the eigenvalues are not degenerate%
\footnote{If the eigenvalues are degenerate, i.e. if the eigenspaces are of dimension greater than~1, then the projection postulate is generalized in the following way: the agent's belief state just after the answer $x_i$ is the normalized projection of her belief state prior to the question \textit{onto the eigenspace corresponding to her answer}. This eigenspace is the ray spanned by $\vert x_i \rangle$ if the eigenvalue is not degenerate, but more generally it can be a plane or a hyperspace. 
Here is why the hypothesis of non-degeneracy is necessary to the reciprocity law. Suppose for simplicity that the Hilbert space is on the real numbers. Suppose that the eigenspace for the eigenvalue 0 for question $A$ is a ray, while the eigenspace for the eigenvalue 0 for $B$ is a plane, with an angle $\pi/4$ between them. Once the answer 0 is obtained from question $A$, the state vector is on the ray, and if $\vert a_0 \rangle$ is projected onto the plane, then there is by hypothesis an angle of $\pi/4$, so $p(b_0|a_0)=\cos^2(\pi/4)$. Now, once the answer 0 is obtained from question $B$, the state vector can be anywhere in the plane, for instance with a right angle to the ray, so $p(a_0|b_0)$ can be null. The reciprocity law does not hold anymore.}. 
It is a quantum law: it is not verified in general by a classical model, in which 
$P(b_j|a_i) = P(a_i|b_j) \times P(b_j)/P(a_i)$, so $P(b_j|a_i) \neq P(a_i|b_j)$ as soon as $P(b_j) \neq P(a_i)$.

For quantum-like models of judgment, the law of reciprocity is somehow well-known\footnote{\label{note_refs-law-of-reciprocity}Cf. 
Franco (2009, 417--418), Wang and Busemeyer (2013, 697--698), Busemeyer \textit{et al.} (2011, 197), Pothos and Busemeyer (2013, 317), Wang \textit{et al.} (2014, 5).}, but it has not been fully investigated.
In the case of our model, it can be instantiated in the following ways:

\begin{numcases}
\strut  
\label{eq_pb0a0=pa0b0}
p(b_0|a_0) = p(a_0|b_0),\\
\label{eq_pb1a0=pa0b1}
p(b_1|a_0) = p(a_0|b_1), \\
\label{eq_pb0a1=pa1b0}
p(b_0|a_1) = p(a_1|b_0),\\
\label{eq_pb1a1=pa1b1}
p(b_1|a_1) = p(a_1|b_1).
\end{numcases}
For each of these equations, the left member is about questions posed in one sense ($A$-then-$B$), while the right member is about the other sense ($B$-then-$A$). This set of equations is actually equivalent to another set, in which each equation is about one order of questions:

\begin{numcases}
\strut  
\label{eq_pb0a0=pb1a1}
p(b_0|a_0) = p(b_1|a_1),\\
\label{eq_pa0b0=pa1b1}
p(a_0|b_0)  = p(a_1|b_1),\\
\label{eq_pb1a0=pb0a1}
p(b_1|a_0) = p(b_0|a_1),\\
\label{eq_pa0b1=pa1b0}
p(a_0|b_1) = p(a_1|b_0).
\end{numcases}

Here's a proof of the first equation (others are similar): 
$p(b_0|a_0) = 1 - p(b_1|a_0)$ because there are only two possible answers ($b_0$ and $b_1$) to question $B$;  then $1 - p(b_1|a_0) = 1 - p(a_0|b_1)$ because of eq.~\ref{eq_pb1a0=pa0b1}; then $1 - p(a_0|b_1) = p(a_1|b_1)$ because there are only two possible answers to question $A$; then $p(a_1|b_1) = p(b_1|a_1)$ because of eq.~\ref{eq_pb1a1=pa1b1}.

This set of equations \ref{eq_pb0a0=pb1a1} to \ref{eq_pa0b1=pa1b0} shows that there exist some equations that a quantum-like model must satisfy \textit{even if the questions are not reversed}, although the reciprocity law (eq.~\ref{eq_law-reciprocity-general}) is originally about questions in reverse orders. The reciprocity law actually gives some constraints on one unchanged experimental setup too, when instantiated for all cases. 
The equivalence of the two sets of equations (reversing-the-order, eq.~\ref{eq_pb0a0=pa0b0}-\ref{eq_pb1a1=pa1b1},  and not-reversing-the-order, eq.~\ref{eq_pb0a0=pb1a1}-\ref{eq_pa0b1=pa1b0}) suggests that there is nothing special about equations which compare reversed orders.

Eq.~\ref{eq_pb0a0=pb1a1} is equal to eq.~\ref{eq_pa0b0=pa1b1}, because of eq.~\ref{eq_pb0a0=pa0b0} or \ref{eq_pb1a1=pa1b1}; similarly for eq.~\ref{eq_pb1a0=pb0a1} and \ref{eq_pa0b1=pa1b0}. So, a new set of equations, that we call the Grand Reciprocity equations, or just the GR equations, can be written:

\begin{numcases}
 \strut  \label{eq_mix0011}
\boxed{p(b_0|a_0) = p(a_0|b_0) = p(b_1|a_1) = p(a_1|b_1),}\\
\label{eq_mix1001}
\boxed{p(b_1|a_0) = p(a_0|b_1) = p(b_0|a_1) = p(a_1|b_0).}
\end{numcases}
This set of equations is equivalent to either of the two previous sets, but it should be preferred as its form is more explicit. It is also equivalent to the reciprocity law itself, because it states it for all possible cases in the model.
Further, eq.~\ref{eq_mix0011} and \ref{eq_mix1001} are equivalent to one another, because  $p(y_0|x_i) + p(y_1|x_i) = 1$ by definition. 
So, each of these equations is actually equivalent to the other sets and to the reciprocity law itself.

To the best of our knowledge, these equations have not yet been fully written in the literature on quantum-like models. Many papers note the law of reciprocity and the consequences for conditional probabilities (cf. footnote~\ref{note_refs-law-of-reciprocity}), but always for one order of the questions only.

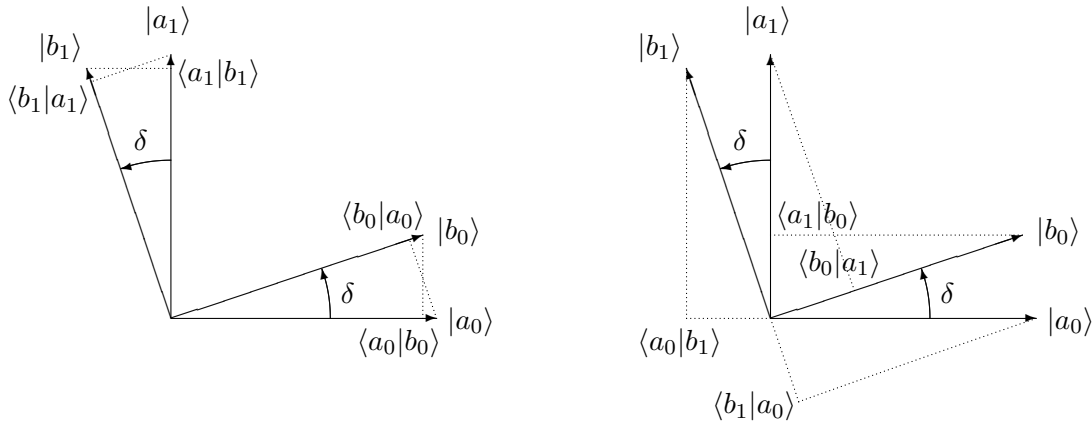
\begin{figure}[!ht]
\begin{center}
\setlength{\unitlength}{.7mm}
\begin{picture}(100,80)(-30,-20) 
\put(0, 0){\vector(1,0){50}}
\put(52, -2){$\vert a_0 \rangle$}
\put(0, 0){\vector(0,1){50}}
\put(-5, 55){$\vert a_1 \rangle$}
\put(0, 0){\vector(3,1){47.4}} 
\put(50, 15){$\vert b_0 \rangle$}
\put(0, 0){\vector(-1,3){15.8}} 
\put(-25, 50){$\vert b_1 \rangle$}
\qbezier(30,0)(30,5)(28.4,9.4)
\put(28.4, 9.4){\vector(-1,4){.01}}
\put(32,3){$\delta$}
\qbezier(0,30)(-5,30)(-9.4,28.4)
\put(-9.4, 28.4){\vector(-4,-1){.01}}
\put(-7,32){$\delta$}
\qbezier[20](47.4,0)(47.4,7.6)(47.4, 15.8)
\put(35,-6){$\langle a_0 \vert b_0 \rangle$}
\qbezier[20](50,0)(47.4,7.6)(45, 14.6) 
\put(32,18){$\langle b_0 \vert a_0 \rangle$}
\qbezier[20](-15.8,47.4)(-7.6,47.4)(0, 47.4)
\put(1,45){$\langle a_1 \vert b_1 \rangle$}
\qbezier[20](-14.6,45)(-7.6,47.4)(0,50) 
\put(-31,40){$\langle b_1 \vert a_1 \rangle$}
\end{picture}
\qquad
\begin{picture}(100,80)(-30,-20) 
\put(0, 0){\vector(1,0){50}}
\put(52, -2){$\vert a_0 \rangle$}
\put(0, 0){\vector(0,1){50}}
\put(-5, 55){$\vert a_1 \rangle$}
\put(0, 0){\vector(3,1){47.4}} 
\put(50, 15){$\vert b_0 \rangle$}
\put(0, 0){\vector(-1,3){15.8}} 
\put(-25, 50){$\vert b_1 \rangle$}
\qbezier(30,0)(30,5)(28.4,9.4)
\put(28.4, 9.4){\vector(-1,4){.01}}
\put(32,3){$\delta$}
\qbezier(0,30)(-5,30)(-9.4,28.4)
\put(-9.4, 28.4){\vector(-4,-1){.01}}
\put(-7,32){$\delta$}
\qbezier[60](47.4,15.8)(23.7,15.8)(0, 15.8)
\put(1,18){$\langle a_1 \vert b_0 \rangle$}
\qbezier[60](0,50)(7.9,27.1)(15.8,5.3)
\put(5,9){$\langle b_0 \vert a_1 \rangle$}
\qbezier[60](-15.8,47.4)(-15.8,23.7)(-15.8,0)
\qbezier[20](-15.8,0)(-7.9,0)(0,0)
\put(-25,-6){$\langle a_0 \vert b_1 \rangle$}
\qbezier[60](50,0)(27.1,-7.9)(5.3,-15.8)
\qbezier[20](5.3,-15.8)(2.7,-7.9)(0,0)
\put(-11,-18){$\langle b_1 \vert a_0 \rangle$}
\end{picture}
\end{center}
\caption{A graphical illustration of the Grand Reciprocity equations, in the special case of a real Hilbert space. A $\delta$ angle is enough to define the two orthonormal bases relatively from one another. Then, conditional probabilities, of the form $\vert \langle x_i \vert y_j \rangle \vert ^2 $, are either equal to $\cos^2(\delta)$ [Left] or to $\sin^2(\delta)$ [Right], with a sum to 1. 
 }\label{fig_relation-bases-A-and-B}
\end{figure}

The GR equations set the value for all possible conditional probabilities: among the eight quantities that can be experimentally measured, there is just one free real parameter. In the case of a real Hilbert space, the origin of this constraint is to be found in the $\delta$ angle between the two bases (cf. fig.~\ref{fig_relation-bases-A-and-B}).

\subsection{Generalizations} 
\label{subsec_generalizations}

Let us now consider some generalizations of the GR equations to cases which go beyond the non-degenerate quantum-like model presented in Section~\ref{sec_general-quantum-model}.

First, the GR equations have been shown for a single agent; by averaging, they obviously still hold for any population of $N$ agents, with various initial state vectors $| \psi_k \rangle$, with $k \in \{1, ..., N\}$.
Another possible generalization concerns the basis vectors: what if they depend on the agent $k$, with 
$\vert a_{0, k} \rangle$, $\vert a_{1, k} \rangle$, $\vert b_{0, k} \rangle$, $\vert b_{1, k} \rangle$?\footnote{This situation is considered for instance by Busemeyer \textit{et al.} 2011 p. 212.}
In this case, the above GR equations can be established for each agent $k$ in the same way as previously, except that the terms are now indexed for each agent $k$, in the form $p_k(x|y)=p_k(z|t)$ (or respectively $=1-p_k(z|t)$), with $(x,y,z,t) \in \{ a_{0,k}, a_{1,k}, b_{0,k}, b_{1,k}\}^4$.
For the whole population, $p(x|y)$ is now defined by the average of the statistical probabilities $p_k$ on all $N$ agents:
\begin{equation}
p(x|y) = 
\frac{1}{N}  \Sigma_k p_k(x|y),
\end{equation}
and similarly for $p(z|t)$. But because for any $k$, $p_k(x|y)=p_k(z|t)$, then
\begin{equation}
 \frac{1}{N}  \Sigma_k p_k(x|y) = \frac{1}{N}  \Sigma_k p_k(z|t).
\end{equation}
So finally $p(x|y) = p(z|t)$ (or respectively = $1-p(z|t)$), and the GR equations hold for the whole population even if the basis vectors depend on the agent. This is just because it holds for any agent, and that the average on the agents is a linear operation that enables to keep the form of the equations.

Also, quantum systems can be described with more general mixed states, instead of pure states. 
Here is the difference: $\vert \psi \rangle$ has been supposed to be equal to a vector from the Hilbert space, i.e. a pure state, and this supposes that the modeler knows to which vector the state is equal. But there are cases in which the modeler does not know exactly to which vector the state is equal (for instance, she only knows that there are 50\% chances that the state is $\vert a_0 \rangle$, and 50\% chances that it is $\vert a_1 \rangle$). To express this, the belief state is considered as a statistical ensemble of pure states, and it is represented by a density matrix (for instance, we write $\rho = 0.5 \vert a_0 \rangle \langle a_0 \vert + 0.5 \vert a_1 \rangle \langle a_1 \vert$ for the above case).
The question is then: are the GR equations valid for pure states only, or also for mixed states? 
For both, because no particular hypothesis has been made on the state \textit{before the first question is asked}. What matters for the demonstration is the state after the first question, and it is doomed to be the corresponding eigenvector (up to a phase factor), whether the belief state was described as a statistical mixture or not before the question was asked. So, the GR equations hold even if mixed states are assumed, which can prove very powerful.

The GR equations apply to a model with 2 questions with 2 possible answers each, but one could also consider a larger set of questions. Consider a model with $q$ questions $A$, $B$, $C$, \dots that are asked in a row, in various orders, each with 2 possible non-degenerate answers, in a Hilbert space of dimension~2. For the questions considered 2 by 2, the reciprocity law (eq.~\ref{eq_law-reciprocity-general}) holds, since what matters is that answers are represented by 1D subspaces, and so the GR equations hold too. This gives a set of  $q \cdot (q-1)/2$ GR equations 
that this model must satisfy. 
Consider now a model with 2 questions, each with $r$ possible answers,
with non-degenerate eigenvalues, in a complex Hilbert space of dimension~$r$. The reciprocity law still holds in this case, and the GR equations hold for each couple of indexes $(i, j) \in \{1, ..., r\}^2$. 
Overall, the GR equations can be generalized to apply for models with $q$ questions and $r$ possible non-degenerate answers.

The hypothesis that eigenvalues are non-degenerate, i.e. that eigenspaces are of dimension~1, can be relaxed in some cases. Suppose a model $M$ with eigenspaces of dimensions greater than 1 is empirically equivalent to a model $M'$ with eigenspaces of dimension~1. So to speak, the supplementary dimensions of $M$ are theoretically useless, and empirically meaningless. 
As the GR equations hold for $M'$, they will also hold for $M$, to which it is equivalent. 
So, more generally, the GR equations hold for models which have eigenspaces of dimension 1, or which are reducible (i.e. equivalent) to it.

These results may have a larger impact still. Suppose that there exists a set of elementary questions, the answers of which would be represented by non-degenerate eigenspaces, i. e. rays. These elementary rays would define the fundamental basis belief states, on which any belief could be decomposed. The number of such elementary rays would define the dimensionality of the Hilbert space representing human beliefs.
Such an assumption is widespread in the quantum-like literature. For instance, a similar idea is expressed by Pothos and Busemeyer about emotions: 
\begin{quote}
{\small
``one-dimensional sub-spaces (called rays) in the vector space would correspond to the most elementary emotions possible. The number of unique elementary emotions and their relation to each other determine the overall dimensionality of the vector space. Also, more general emotions, such as happiness, would be represented by subspaces of higher dimensionality." (Pothos and Busemeyer 2013, 258). }
\end{quote} 
Generalized GR equations would hold for such elementary questions. Then, all models with degenerate eigenspaces would just be coarse-grained models, which could be refined with more elementary questions and rays. For instance, a degenerate answer represented by a plane combines two elementary dimensions because it fails to distinguish between them. As a consequence, the meaning of a degenerate answer could be specified through more specific elementary answers. More importantly, the degenerate model could actually be tested on these fundamental questions, with our GR equations or their generalized versions with $q$ questions and $r$ answers. In other words, if elementary questions existed --- and it seems to be a usual assumption in the quantum-like literature
--- any model of the kind presented in Section~2, but with non-degenerate answers or not, could be tested with GR equations. This is of fantastic interest, as it enables to test any model, and it has never been noted before. We shall come back on this point in Section~6.

\subsection{Link with double stochasticity}
\label{subsec_link-dble-stoch}

Part of the GR equations is actually known in the literature as the requirement of ``double stochasticity".
Let us analyze the links between the two constraints.

Define the change of basis matrix $\mu^{a,b}$ between the two bases $(\vert a_0 \rangle, \vert a_1 \rangle)$ and $(\vert b_0 \rangle, \vert b_1 \rangle)$, as:

\begin{equation}
\mu^{a,b} = 
\begin{pmatrix}
\langle a_0 \vert b_0 \rangle & \langle a_0 \vert b_1 \rangle \\
\langle a_1 \vert b_0 \rangle & \langle a_1 \vert b_1 \rangle
\end{pmatrix}.
\end{equation}
As the two bases are orthonormal by hypothesis, this matrix is unitary.
From it, a transition matrix $T^{a, b}$ can be defined by $T^{a,b}_{ij} = |\mu^{a,b}_{ij}|^2$:
\begin{equation} \label{eq_Tab}
T^{a,b} = 
\begin{pmatrix}
\vert \langle a_0 \vert b_0 \rangle\vert^2 & \vert \langle a_0 \vert b_1 \rangle \vert^2 \\
\vert\langle a_1 \vert b_0 \rangle \vert^2& \vert \langle a_1 \vert b_1 \rangle \vert^2
\end{pmatrix}
=
\begin{pmatrix}
p(a_0|b_0) & p(a_0|b_1) \\
p(a_1|b_0) & p(a_1|b_1)
\end{pmatrix}.
\end{equation}
$T^{a,b}$ contains the probabilities for the answers to question $A$ given the previous answer to question $B$ ($T^{a,b}$ could be also called $T^{A|B}$).

Transition matrices are left stochastic: they are square matrices of non-negative real numbers, of which each column sums to 1. This expresses the fact that, once an answer has been given to the first question, there is a probability 1 that one of the answers is given to the second question.
A matrix is said to be doubly stochastic in case all columns and all rows sum to~1. 
Saying that $T^{a,b}$ is doubly stochasticity amounts to the following equations:
\begin{numcases}
\strut 
\label{eq_pa0b0+pa1b0=1}
p(a_0|b_0) + p(a_1|b_0) = 1\\
p(a_0|b_1) +p(a_1|b_1) = 1\\
p(a_0|b_0) + p(a_0|b_1) = 1\\
\label{eq_pa1b0+pa1b1=1}
p(a_1|b_0) + p(a_1|b_1) = 1
\end{numcases}
Subtracting these equations one by one gives eq.~\ref{eq_pa0b0=pa1b1} and \ref{eq_pa0b1=pa1b0} 
--- but no more\footnote{Eq.~\ref{eq_pb0a0=pb1a1} and \ref{eq_pb1a0=pb0a1} cannot be derived from eq.~\ref{eq_pa0b0+pa1b0=1}--\ref{eq_pa1b0+pa1b1=1}: as they bear on questions asked in different orders, it is easy to imagine a case in which the former are false, whereas the latter are true.}.
So, one gets \textit{one half} of the reciprocity relations. This should be no surprise: the double stochasticity constraint bears on only \textit{one} experiment, when the order of the questions is the same, and cannot be informative about the reverse order. 
It is easy to see that if the double stochasticity of the reverse transition matrix ($T^{b,a}$) is assumed, then one gets the other half of our reciprocity relations (eq.~\ref{eq_pb0a0=pb1a1} and \ref{eq_pb1a0=pb0a1}), and then also the GR equations. Conversely, our GR equations imply double stochasticity for both transition matrices. So, the GR equations are equivalent to the requirement of double stochasticity for transition matrices in both senses.

In the literature, double stochasticity is a well-known property of transition matrices for quantum-like non-degenerate models\footnote{\label{note_refs-double-stoch}
Cf. for instance Busemeyer and Bruza (2012, 53--54), Busemeyer, Wang and Lambert-Mogiliansky (2009), 
Pothos and Busemeyer (2013, 269).}, 
but it has generally been required for \textit{one} transition matrix only (i.e. for questions posed in one order). 
An exception is Khrennikov (2010, 24 and 36), who studies double stochasticity for transition matrices in both senses, and shows that it must be verified by quantum-like probability models with non-degenerate eigenvalues, i.e. like in our general model. However, he does not insist on testing experimentally this property in a systematic way, as we shall do here, with important consequences for existing models.

Although the GR equations are equivalent to double stochasticity in both orders for transition matrices, there are several reasons why the former formulation should be preferred, both pragmatically and theoretically.
To test double stochasticity, one first needs to write equations like eq.~\ref{eq_pa0b0+pa1b0=1}--\ref{eq_pa1b0+pa1b1=1}. But such equations are not independent, so it complicates the statistical tests of significance. Or in order not to test useless equations, some work needs to be done beforehand to find the independent equations. Actually, untangling the equations would just lead to the GR equations, which are directly usable. Overall, the GR equations can be considered as the pragmatic form one should use to test double stochasticity in both orders. They also clearly show that two set of equations (\ref{eq_mix0011} and \ref{eq_mix1001}) are equivalent, and that one needs to pick only the corresponding data. 
Also, the GR equations show at first sight that there is only one free parameter among all the possible conditional probabilities, while this is not transparent from the double stochasticity requirement. 
More theoretically, the GR equations are directly linked with a central mathematical property of the scalar product in the model, making it clear why it holds. On the other hand, the reason why double stochasticity should hold in this model is more obscure. Finally, the GR equations connect with a fundamental law of quantum mechanics, namely the law of reciprocity, which helps to understand why it holds only in the case of non-degenerate eigenvalues, like in quantum mechanics.

\subsection{Link with the QQ equality}
\label{subsec_link-QQ-equality}

Wang and Busemeyer (2013) and Wang \textit{et al.} (2014) defend the test of a relation called ``the QQ equality" that they introduce. This equality holds for the non-degenerate quantum-like model presented in Section~\ref{sec_general-quantum-model}, and also if eigenvalues were degenerate. In the case of non-degenerate eigenvalues, it can be derived from our GR equations, as can be shown in the following way.

With the notations of our model (and adding the notation $p_{XY}(z_i)$ for the probability to answer $i$ to question $Z$, when posed in the order $X$-then-$Y$), the QQ equality (Wang and Busemeyer 2013, 698) can be written as:
\begin{equation}
p(a_0, b_1) + p(a_1, b_0) = p(b_0, a_1) + p(b_1, a_0),
\end{equation}
that is,
\begin{equation}
p(b_1|a_0)  p_{AB}(a_0) + p(b_0|a_1)  p_{AB}(a_1) = p(a_1|b_0)  p_{BA}(b_0) + p(a_0|b_1)  p_{BA}(b_1).
\end{equation}
Because $p_{XY}(z_0) + p_{XY}(z_1) = 1$, it is equivalent to:
\begin{equation}
p(b_1|a_0)  p_{AB}(a_0) + p(b_0|a_1)  [1-p_{AB}(a_0)] = p(a_1|b_0)  p_{BA}(b_0) + p(a_0|b_1)  [1- p_{BA}(b_0)],
\end{equation}
which can be rewritten as:
\begin{equation}
[p(b_1|a_0) - p(b_0|a_1)] p_{AB}(a_0) + p(b_0|a_1)  = [p(a_1|b_0) - p(a_0|b_1)]  p_{BA}(b_0) + p(a_0|b_1).
\end{equation}
The square brackets are null because of eq.~\ref{eq_pb1a0=pb0a1} and \ref{eq_pa0b1=pa1b0}, and the two remaining terms are equal because of eq.~\ref{eq_mix1001}. So, the QQ equality is demonstrated from the GR equations when eigenspaces are not degenerate.

The GR equations cannot be derived in general from the QQ equality: below, Section~\ref{sec_empirical-glimpse} gives an example of data which satisfy the QQ equality, but not our GR equations. So the two tests are not equivalent. 
They test different things: if a data set does not satisfy the GR equations, the non-degenerate quantum-like model of Section~2 cannot apply to it (it remains an open question whether a degenerate model would); 
if a data set does not satisfy the QQ equality, the non-degenerate quantum-like model of Section~2 cannot apply to it, and a degenerate version of it could not either. In other words, a non-degenerate quantum-like model must satisfy both the GR equations and the QQ equality, and a degenerate quantum-like model must satisfy the QQ equality.

\subsection{Testing the Grand Reciprocity equations}
\label{subsec_testing-GRE}

Why test these GR equations?
As noted before, the quantum-like literature often consider models that can be either non-degenerate (answers are represented by rays) or degenerate (answers are represented by subspaces of any dimension).
So, a first reason to test the GR equations is simplicity: if a non-degenerate model is empirically adequate, why bother considering a degenerate one? The possibility to have a non-degenerate model should be settled, and the GR test exactly enables it.
Conversely, for models that are proposed only in a non-degenerate version, the GR equations should be verified without condition, and they provide a decisive test of empirical adequacy.
So, in any case, we suggest that (one of) the GR equation(s) be directly tested experimentally for any quantum-like model that contains the model presented in Section~2, as a particular case only or not.
``Directly tested" here means that what should be measured are the conditional probabilities that are present in these equations, or data that enable their computation.

On a more pragmatic ground, 
the needed data are basic ones, easy to get experimentally. 
The equations are ready-to-use ones, simpler than the test of double stochasticity (cf. Section~\ref{subsec_link-dble-stoch}).
More theoretically, each GR equation is equivalent to a fundamental property of quantum-like models, namely the law of reciprocity. It is not just a consequence of it 
but it is just the expression of this law \textit{for all possible cases} in the model. Testing the GR equations is exactly testing the reciprocity law (testing the QQ equality, which is only a consequence of the reciprocity law, amounts to another test). The benefit is that testing the GR equations is more economical: they state that only 4 conditional probabilities need to be compared to test the law in general.
Another reason to test the GR equations is that a classical model does not verify them; so they can be seen as a test of the quantum-like character of the data.\footnote{Indeed, the GR equations are equivalent to the reciprocity law, which is not true in general for classical models (cf.  Section~\ref{subsec_derivation-recipro-relations}). Note that classical models which satisfy the GR equations are uninteresting: all $P(x_i)$ and all $P(x_i|y_j)$ are equal to $0.5$.}

A last point in favor of the test is that, when some empirical data verify the GR equations, it can be used to define a quantum-like model in a real Hilbert space.
Define $p =p(a_0|b_0)$; because of the GR equations, any conditional probability $p(x_i|y_j)$ is either $p$ or $1-p$, and eq.~\ref{eq_Tab} becomes:
\begin{equation}
T^{a,b} = \begin{pmatrix}
p & 1-p  \\
1-p & p
\end{pmatrix}.
\end{equation}
On the other hand, a model from a real Hilbert space (like in fig.~\ref{fig_bases-A-and-B}) has a change of basis matrix
\begin{equation}
\mu^{a,b} = \begin{pmatrix}
\cos \delta & - \sin \delta  \\
\sin \delta & \cos \delta
\end{pmatrix},
\end{equation}
and hence a transition matrix
\begin{equation}
T^{a,b} = \begin{pmatrix}
\cos^2 \delta & \sin^2 \delta  \\
\sin^2 \delta & \cos^2 \delta
\end{pmatrix}.
\end{equation}
Taking $\delta = \arccos(\sqrt{p})$ enables the real model to account for the data of conditional probabilities. 
To put it another way, checking the empirically validity of the GR equations enables at the same time to define the two bases relatively from one another.

Note that testing the GR equations require that the experiment be conducted in both question orders ($A$-then-$B$ and $B$-then-$A$). If only one question order is studied, a test can still be performed, but only on half of the equations. 
So, whenever a quantum-like model of the kind depicted here is proposed to account for the results of a succession of two questions in one order only, then the model should also be tested for the reverse order.

\section{Applying the GR equations to non-degenerate (versions of) existing models}
\label{sec_test-order-effects}

The non-degenerate quantum-like model that we introduced in Section~\ref{sec_general-quantum-model} was intended to serve as a concise presentation for several models from the literature.
In this section, we discuss in detail the relation between our model and existing ones, in order to clarify whether the latter should verify the GR equations. Recall that satisfying the GR equations is required in general for non-degenerate models only. So, if an existing model is proposed in the most general case either as non-degenerate or degenerate, then the GR equations apply only to the non-degenerate version of the model, not to its degenerate version.

Consider first Conte \textit{et al.} (2009), who propose a quantum-like model to account for order effects in mental states during visual perception of ambiguous figures.
Their model can be cast into the lines of our general model: the two dichotomous questions or tests are also called $A$ and $B$, and the probabilities of the answers are noted ${p(A=+)}$, ${p(A=-)}$, ${p(B=+)}$ and ${p(B=+)}$, corresponding to our $p(a_0)$, $p(a_1)$, $p(b_0)$ and $p(b_1)$. The questions concern visual perception of ambiguous figures, and a vector state, noted here $\phi$ (instead of our $\psi$) represents the state of consciousness about the perception. It belongs to a complex Hilbert space, and an answer is represented by a one-dimensional subspace (this is implicit in their formula (3)). 
The usual Born rule makes the link with probabilities. 
The model also involves a projection postulate: during perception, the ``potential" state of consciousness is collapsed onto ``an actual or manifest state of consciousness" (p.~6 and 7). 
Note however a slight difference: for Conte \textit{et al.}, the projection of the state arises during perception of the figure, not during answering a question. But as the latter quickly follows the former, and as an agent answers a question only once her perception has stabilized, this does not make any difference in practice. 
Overall, Conte \textit{et al.} (2009)'s model matches our general model in all respects, as it supposes that eigenspaces are of dimension~1. So, it has to obey the GR equations.

\bigskip

Consider now the model proposed by Wang and Busemeyer (2013), called the QQ model, which is to account for several types of order effects about judgments on various topics (politics, sports, society, etc.). The clear presentation of the model in their Section~2, together with their geometric approach which provides figures similar to ours (in the case of a real Hilbert space), easily enables to see that it encompasses the non-degenerate model of Section~2. However, Wang and Busemeyer's model is more general and allows for an answer to be represented either with a subspace of dimension~1 (non-degenerate version), or with a subspace of dimension larger than~1 (degenerate version). The GR equations apply only to the former version, as a special case. For instance, they apply to figures 1 and 2 (p.~4 and 5), which represent ``a two-dimensional example of the quantum-like model of question order effects".
Hence, the GR test will exactly fit here the aim we set it in the introduction: within this general model, specify whether non-degenerate model can be sufficient, or degenerate model are required.

Wang and Busemeyer's general model is also at the basis of two other presentations: in Busemeyer and Bruza's 2012 book (p.~99--116), and in the review article of Pothos and Busemeyer (2013). 
Here too, the model can be non-degenerate or degenerate, and the GR equations apply with the same conditions: only to the special case of the non-degenerate version.

Atmanspacher and Römer (2012, p.~277) make the suggestion that Wang and Busemeyer's model could be extended to mixed states, instead of just pure states. Although this is an interesting theoretical possibility, it is clear that it cannot be used to avoid the GR equations, as they are valid for pure as for mixed states (cf. Section~\ref{subsec_generalizations}).

Building on the paper of Wang and Busemeyer (2013), Wang \textit{et al.} (2014) take up the same quantum-like model for order effect, in order to apply it to a much larger set of experiments. Here again, answers are supposed to be represented with subspaces of any dimension. So, the GR equations apply only to the special case of the model in its non-degenerate version, i. e. when answers are represented with rays as in the model of Section~2.

\section{An empirical glimpse: the GR tests}
\label{sec_empirical-glimpse}

The previous section has listed several cases in which the GR equations can be tested. For all cases but Conte \textit{et al.} (2009), the models have been proposed for any $N$ dimensions, i. e. as either non-degenerate or degenerate. Then, the GR equations can test the special case of the non-degenerate version of these models, in order to determine whether this simplest case is sufficient, or if the more general degenerate models are required.
This section is concerned with performing these empirical GR tests.
They are done using the empirical data that are provided in the papers, 
and which come from either laboratory or field experiments. 
To study these empirical data, we divide them into two sets, according to the topic and the authors: first, in Sections \ref{data set} to \ref{statRes}, we consider the data on experiments about judgment on social or political questions, which enable to test the model first proposed by Wang and Busemeyer (2013) and considered by 
Busemeyer and Bruza (2012), Pothos and Busemeyer (2013), and Wang \textit{et al.} (2014); second, in Section~\ref{subsec_res-conte}, the data concern an experiment about visual perception, and we test the model proposed by Conte \textit{et al.} (2009). 

\subsection{First data set} \label{data set}
Our first data set 
 gathers 72 experiments on order effects, with 70 national surveys made in the USA plus 2 laboratory experiments: 
\begin{itemize}
\item[--] 66 national surveys from the Pew Research Center run between 2001 and 2011 about politics, religion, economic policy and so on, considered in Wang \textit{et al.} (2014) (noted in Table~\ref{tab_questions-data-set} as experiments 1 to 66);
\item[--] 3 surveys from Moore (2002) concerning Gallup public opinion polls (noted as experiments 67 to 69), 
which are considered in Wang and Busemeyer (2013) and in Wang \textit{et al.} (2014) ; 
\item[--] 2 laboratory experiments conducted by Wang and Busemeyer (2013). 
In the first experiment, the authors replicate in a laboratory setting the experiment proposed by Moore (2002) about racial hostility, whereas, in the second one, they replicate the survey of Wilson \textit{et al.} (2008) (experiments 70 and 71) ;
\item[--] one further Gallup survey reported by Schuman and Presser (1981), also considered in Wang \textit{et al.} (2014) 
(experiment 72).
\end{itemize}

\begin{table}[tb]
\caption{Questions of the experiments included in the first data set. 
\label{tab_questions-data-set}}
\begin{tabular}{ >{}m{1cm} >{}m{0.4cm} l }
 \multicolumn{3}{ l }{1-26 (26 experiments).} \\ \hline 
& A. & Do you approve or disapprove of the way Bush/Obama is handling\\
& & the job as President?\\
& B. & All in all, are you satisfied or dissatisfied with the way things are going\\
& & in this country today?\\
 \multicolumn{3}{ l }{27-41 (15 experiments).}\\ \hline 
& A. & Do you approve or disapprove of the job the Republican leaders \\
&& in Congress are doing?\\
& B. & Do you approve or disapprove of the job the Democratic leaders \\
& & in Congress are doing?\\
 \multicolumn{3}{ l }{42-66 (25 experiments).}\\ \hline 
&  \multicolumn{2}{ l }{These experiments include diverse questions covering topics, from religious}\\
& \multicolumn{2}{ l }{beliefs to support to economic policy.}\\
 \multicolumn{2}{ l }{67.} & \\ \hline
& A. & Do you generally think Bill Clinton is honest and trustworthy?\\
& B. & Do you generally think Al Gore is honest and trustworthy?\\
 \multicolumn{2}{ l }{68.} & \\ \hline
& A. & Do you think Newt Gingrich is honest and trustworthy?\\
& B. & Do you think Bob Dole is honest and trustworthy?\\
 \multicolumn{2}{ l }{69 and 70.} & \\ \hline
& A. & Do you think that only a few or many white people dislike black people?\\
& B. & Do you think that only a few or many black people dislike white people?\\
 \multicolumn{3}{ l }{71.} \\ \hline 
& A. & Do you generally favor or oppose affirmative action (AA) programs for \\
& & racial minorities?\\
& B. & Do you generally favor or oppose affirmative action (AA) programs for \\
& & women?\\
 \multicolumn{2}{ l }{72.} & \\ \hline 
& A. & Do you think it should be possible for a pregnant woman to obtain\\
& &  a legal abortion if she is married and does not want any more children?\\
& B. & Do you think it should be possible for a pregnant woman to obtain\\
& & a legal abortion if there is a strong chance of serious defect in the baby?\\

\end{tabular}
\end{table}

Figure \ref{boxplot} reports the box plot of the distribution of the total number of subjects per experiment. The median experiment (741 subjects) is thus supposed to involve half of the samples (about 370 subjects) in the $A$-then-$B$ treatment and the remaining half in the $B$-then-$A$ treatment. The circle on the left part of the box-plot, outside the left whisker, pinpoints the two laboratory experiments 70 and 71, which involve the smallest sample, that is, 228 subjects.
\begin{figure}[tb]
\begin{tikzpicture}[x=1pt,y=1pt, trim left= 1.3cm ]
\pgfresetboundingbox
\definecolor[named]{fillColor}{rgb}{1.00,1.00,1.00}
\path[use as bounding box,fill=fillColor,fill opacity=0.00] (0,0) rectangle (505.89,108.41);
\begin{scope}
\path[clip] ( 60.00, 36.00) rectangle (445.89, 72.40);
\definecolor[named]{drawColor}{rgb}{0.00,0.00,0.00}

\path[draw=drawColor,line width= 1.2pt,line join=round] (206.67, 47.46) -- (206.67, 60.94);

\path[draw=drawColor,line width= 0.4pt,dash pattern=on 4pt off 4pt ,line join=round,line cap=round] (119.31, 54.20) -- (180.86, 54.20);

\path[draw=drawColor,line width= 0.4pt,dash pattern=on 4pt off 4pt ,line join=round,line cap=round] (282.60, 54.20) -- (224.09, 54.20);

\path[draw=drawColor,line width= 0.4pt,line join=round,line cap=round] (119.31, 50.83) -- (119.31, 57.57);

\path[draw=drawColor,line width= 0.4pt,line join=round,line cap=round] (282.60, 50.83) -- (282.60, 57.57);

\path[draw=drawColor,line width= 0.4pt,line join=round,line cap=round] (180.86, 47.46) --
	(180.86, 60.94) --
	(224.09, 60.94) --
	(224.09, 47.46) --
	(180.86, 47.46);

\path[draw=drawColor,line width= 0.4pt,line join=round,line cap=round] (359.78, 54.20) circle (  2.25);

\path[draw=drawColor,line width= 0.4pt,line join=round,line cap=round] (399.26, 54.20) circle (  2.25);

\path[draw=drawColor,line width= 0.4pt,line join=round,line cap=round] (114.31, 54.20) circle (  2.25);

\path[draw=drawColor,line width= 0.4pt,line join=round,line cap=round] (114.31, 54.20) circle (  2.25);
\end{scope}
\begin{scope}
\path[clip] (  0.00,  0.00) rectangle (505.89,108.41);
\definecolor[named]{drawColor}{rgb}{0.00,0.00,0.00}

\path[draw=drawColor,line width= 0.4pt,line join=round,line cap=round] ( 74.29, 36.00) -- (431.60, 36.00);

\path[draw=drawColor,line width= 0.4pt,line join=round,line cap=round] ( 74.29, 36.00) -- ( 74.29, 30.00);

\path[draw=drawColor,line width= 0.4pt,line join=round,line cap=round] (163.62, 36.00) -- (163.62, 30.00);

\path[draw=drawColor,line width= 0.4pt,line join=round,line cap=round] (252.94, 36.00) -- (252.94, 30.00);

\path[draw=drawColor,line width= 0.4pt,line join=round,line cap=round] (342.27, 36.00) -- (342.27, 30.00);

\path[draw=drawColor,line width= 0.4pt,line join=round,line cap=round] (431.60, 36.00) -- (431.60, 30.00);

\node[text=drawColor,anchor=base,inner sep=0pt, outer sep=0pt, scale=  1.00] at ( 74.29, 14.40) {0};

\node[text=drawColor,anchor=base,inner sep=0pt, outer sep=0pt, scale=  1.00] at (163.62, 14.40) {500};

\node[text=drawColor,anchor=base,inner sep=0pt, outer sep=0pt, scale=  1.00] at (252.94, 14.40) {1000};

\node[text=drawColor,anchor=base,inner sep=0pt, outer sep=0pt, scale=  1.00] at (342.27, 14.40) {1500};

\node[text=drawColor,anchor=base,inner sep=0pt, outer sep=0pt, scale=  1.00] at (431.60, 14.40) {2000};
\end{scope}
\begin{scope}
\path[clip] (  0.00,  0.00) rectangle (505.89,108.41);
\definecolor[named]{drawColor}{rgb}{0.00,0.00,0.00}

\path[draw=drawColor,line width= 0.4pt,line join=round,line cap=round] ( 60.00, 36.00) --
	(445.89, 36.00) --
	(445.89, 72.40) --
	( 60.00, 72.40) --
	( 60.00, 36.00);
\end{scope}
\end{tikzpicture}
\caption{Distribution of number of subjects per experiment.}
\label{boxplot}
\end{figure}
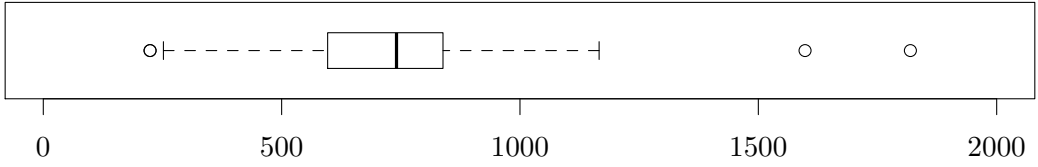\\

Each experiment can be described according to the following common model and notations: two response categorical variables,
\begin{itemize}
\item[--] the Bernoulli random variable $ \mathbf{A} \in \{ a_0, a_1 \}$ that represents the possible replies (respectively ``yes" or ``no") to the dichotomous question $A$,
\item[--] the Bernoulli random variable $ \mathbf{B} \in \{ b_0, b_1\} $ that represents the possible replies (respectively ``yes" or ``no") to the dichotomous question $B$, 
\end{itemize}
and one explanatory categorical variables,
\begin{itemize}
\item[--] the variable $ \mathbf{O} \in \{ 0, 1\}$ that represents the order of appearance of the questions ---  0 for the $A$-then-$B$ order and 1 for the $B$-then-$A$ order.
\end{itemize}
Empirically, what is observed is not conditional probabilities like $p(b_j|a_i)$, but joint frequencies $n(a_i,b_j)$, that is, the outcome of the counting process of the people responding $i$ to the first question $A$ and then $j$ to the second question $B$ (the order matters in this notation: 
$n(b_j, a_i)$  refers to the $B$-then-$A$ experiment). This information can be reported as a contingency table (Table~\ref{ContTab}). 
\begin{table}[h!] 
\caption{The contingency table of a generic experiment.\label{ContTab}}
\begin{center}
  \begin{tabular}{  >{\centering}m{1.45cm} | >{\centering}m{1.8cm} | >{\centering}m{1.8cm} | >{\centering}m{1.8cm} | c |}
    & \multicolumn{2}{ c | }{$\mathbf{O} = 0$} & \multicolumn{2}{ c | }{$\mathbf{O} = 1$} \\ \cline{2-5}
    & $\mathbf{B}=b_0$ & $\mathbf{B}=b_1$  & $\mathbf{B}=b_0$ & $\mathbf{B}=b_1$ \\ \hline
    $\mathbf{A}=a_0$ & $n(a_0, b_0)$ & $n(a_0, b_1)$ & $n(b_0, a_0)$ & $n(b_1, a_0)$  \\ \hline
    $\mathbf{A}=a_1$ & $n(a_1, b_0)$ & $n(a_1, b_1)$ & $n(b_0, a_1)$ & $n(b_1, a_1)$  \\ \hline
  \end{tabular}
\end{center}
\end{table}

Validity conditions for hypothesis testing are sensitive to the number of counts --- it is commonly required at least 5 counts for the joint frequencies. To this end, Table~\ref{descrStatJF} reports some statistical information about the distributions of the joint frequencies over the 72 experiments. In particular, Table \ref{descrStatJF} focuses on the region of the distributions corresponding to small values: the rows report the first decile of each distribution (D1 - 10\%), the smallest number of subjects occurred in one experiment (min) and the number of experiments presenting a joint frequency value lower than 5. Summarizing, more than 90\% of the experiments have joint frequency values greater or equal to 5, 
5 out of 8 joint frequencies never present values less than 5, 
and only 6 experiments present for only one joint frequency a value smaller than 5. 

\begin{table}[h!] 
\caption{Descriptive statistics for the joint frequencies.  \label{descrStatJF}}
\begin{center}
\begin{tabular}{ c | c c c c | c c c c }
 & $n(a_0, b_0)$ & $n(a_0, b_1)$ & $n(a_1, b_0)$ & $n(a_1, b_1)$ & $n(b_0, a_0)$ & $n(b_0, a_1)$ & $n(b_1, a_0)$ & $n(b_1, a_1)$\\ \hline
D1 & 33 & 19 & 5 & 47 & 30 & 9 & 17 & 47 \\
min & 6 & 5 & 2 & 19 & 11 & 2 & 3 & 14\\
n<5 & 0 & 0 & 4 & 0 & 0 & 1 & 1 & 0 \\ \hline
\end{tabular}
\end{center}
\end{table}

\subsection{The statistical test} \label{stTest}

The aim of our statistical study is to study the validity of quantum-like modeling on order effect experiments by testing the theoretical GR equations (eq.~\ref{eq_mix0011} and \ref{eq_mix1001}). As they are equivalent,  we focus on the first set of equations,
\begin{equation}\label{eq_GRtested}
p(b_0|a_0) = p(a_0|b_0) = p(b_1|a_1) = p(a_1|b_1).\\
\end{equation}
The conditional probabilities, which are not observable, are estimated by considering conditional relative frequencies $f(a_i|b_j)$ and $f(b_j|a_i)$, under the hypothesis that relative frequencies converge to probabilities when the number of people involved in the experiment tends to infinity. Eq.~\ref{eq_GRtested} is equivalent to 6 two-by-two comparisons:
\begin{numcases}
\strut 
\label{6Tests}
\text{T1: } \quad f(b_0|a_0) = f(a_0|b_0),\label{firstTest}\\
\text{T2: } \quad f(b_0|a_0) = f(b_1|a_1),\\
\text{T3: } \quad f(b_0|a_0) = f(a_1|b_1),\\
\text{T4: } \quad f(a_0|b_0) = f(b_1|a_1),\\
\text{T5: } \quad f(a_0|b_0) = f(a_1|b_1),\\
\text{T6: } \quad f(b_1|a_1) = f(a_1|b_1).\label{lastTest}
\end{numcases}

Each test is to compare two conditional relative frequencies $y$ and $x$, with the null hypothesis that they are equal.\footnote{\label{ftn_arbitrary} Note that there is some arbitrariness in the comparison of the number of a test between two experiments, as it depends on our labeling of both ``a/b" and ``0/1". For instance, an experiment which fails T1 would fail T6 if it were labeled differently, and similarly for T2 with T5, and T3 with T4.}
From a statistical viewpoint, it is convenient to perform an equivalent test in terms of logits: 
$\log\left(\frac{y}{1-y}\right)=\log\left(\frac{x}{1-x}\right)$, or even whether the natural logarithm of the odds ratio is equal to zero: $\log ({\rm OR}) = \log\left(\frac{x(1-y)}{(1-x)y}\right) = 0$. The sampling distribution of the log odds ratio $X$, under the hypothesis of the large sample approximation, is approximately normal: $X\ \sim\ \mathcal{N}(\log ({\rm OR}),\,\sigma^2).$ 
Instead of conditional relative frequencies, the log odds ratio can be easily expressed equivalently in terms of joint frequencies (cf. Appendix \ref{AppA}).
For instance, instead of computing the first statistical test (eq.~\ref{firstTest}), we can perform the following test:
\begin{equation}
\log ({\rm OR}) = \log\left(\frac{n(a_0, b_0)n(b_0, a_1)}{n(a_0, b_1)n(b_0, a_0)}\right) = 0.
\end{equation}
A continuity correction is also applied, because the normal approximation to the binomial is used, which is effective in particular for small values of $n(a_i, b_j)$ or $n(b_j, a_i)$. 
The above test becomes:
\begin{equation}
\log ({\rm OR}) = \log\left(\frac{(n(a_0, b_0) + 0.5)(n(b_0, a_1) + 0.5)}{(n(a_0, b_1) + 0.5)(n(b_0, a_0) + 0.5)}\right) = 0.
\end{equation}
We suppose here that 
\begin{equation}
\frac{\log ({\rm OR})}{\rm SE_{logOR}} \sim \mathcal{N}(0, 1),
\end{equation}
where ${\rm SE_{logOR}}$ is the standard error of the log odds ratio. It is estimated as the square root of the sum of the inverse of all the joint frequencies that are considered in the estimation of the ${\rm OR}$:\\
\begin{equation}
{{\rm SE_{logOR}} = \sqrt{\dfrac{1}{n(a_0, b_0)} + \dfrac{1}{n(b_0, a_1)} + \dfrac{1}{n(a_0, b_1)} + \dfrac{1}{n(b_0, a_0)}}}.
\end{equation}
Finally, a Bonferroni correction of the type~I error is adopted because of the multiple comparisons (six). Note that this statistical correction is a very conservative one, which means that it makes false positive rejections much less liable to occur. Concretely, the two-tailed test implies the null hypothesis of equality between the two conditional frequencies at the $K\%$ significance level is rejected if:
\begin{equation}
p\text{-value} = 2\cdot\left(1-{\rm CDF_{stdNorm}}\left(\left|\frac{\log ({\rm OR})}{\rm SE_{logOR}}\right|\right)\right)\leq \frac{K}{100}
\end{equation}
where ${\rm CDF_{stdNorm}}$ is the cumulative distribution function of the standard normal distribution (mean = 0 and standard deviation = 1).
Finally, we test by estimating the adjusted $p$-value based on the Bonferroni correction, with
\begin{equation}
\mbox{adjusted $p$-value} = 6 \cdot p\text{-value}.
\end{equation}

\subsection{Results on the first data set} \label{statRes}

Table~\ref{longtableres} in Appendix~\ref{sec_appendix_results} reports the adjusted $p$-values of the tests (columns) for each of the 72 experiments (rows). 
Recall that only one rejection of the six tests T1-T6 is sufficient for a non-degenerate quantum-like model to be considered as empirically inadequate. 
In this sense, satisfying 5 tests over 6 is not more desirable than satisfying 1 --- this is an all-or-nothing problem, or a ``zero-or-non-zero" rejection matter. 

First, let us take a general viewpoint on the rejection question. 
Some summary statistics are provided in Table~\ref{descrStat} relative to the distribution of number of test rejections per experiment. 
The rejection rate per experiment is high, above 3. The distribution of the test rejections is highly skewed with respect to a uniform distribution. More than 75\% of the experiments (Q1 - first quartile) reject half or more than half of the 6 tests and both the median and the mode are equal to 4. The key point is that the first decile (D1) is equal to 1, that is, more than 90\% of the experiments exhibit at least one test rejection, thus confirming the hypothesis that quantum-like modeling in which one dimensional subspaces (rays) represent answers to questions is not empirically adequate for the experiments considered in this first data set.

\begin{table}[tb] 
\caption{Descriptive statistics for the number of test rejections 
per experiment.
\label{descrStat}}
\begin{center}
\begin{tabular}{ l | c }
mean &  3.63 \\
mode & 4 \\
Q3 (75\%) & 5 \\
median & 4 \\
Q1 (25\%) & 3 \\
D1 (10\%) & 1 \\
\end{tabular} 
\end{center}
\end{table}

Second, it is possible to have an individual analysis of the experiments:
as the last column of Table~\ref{longtableres} (Appendix~\ref{sec_appendix_results}) reports the experiment identifier (exp ID) adopted in Table~\ref{tab_questions-data-set}, it is possible to track each question in the ranked-based list of experiments.
For each experimental protocol (laboratory or field experiments, Gallup or Pew Research Center surveys and different authors), at least one experiment rejects at least one test, which shows that  the result of non-satisfaction is robust also against experimental settings and questions. For instance, the two laboratory experiments (70 and 71) of the data set exhibit one test rejection --- as they involve the smallest subject pools, one can conjecture that a higher number of test rejections would occur with more subjects involved. 

\begin{figure}[!tb]
\begin{center}
\includegraphics[scale=0.6]{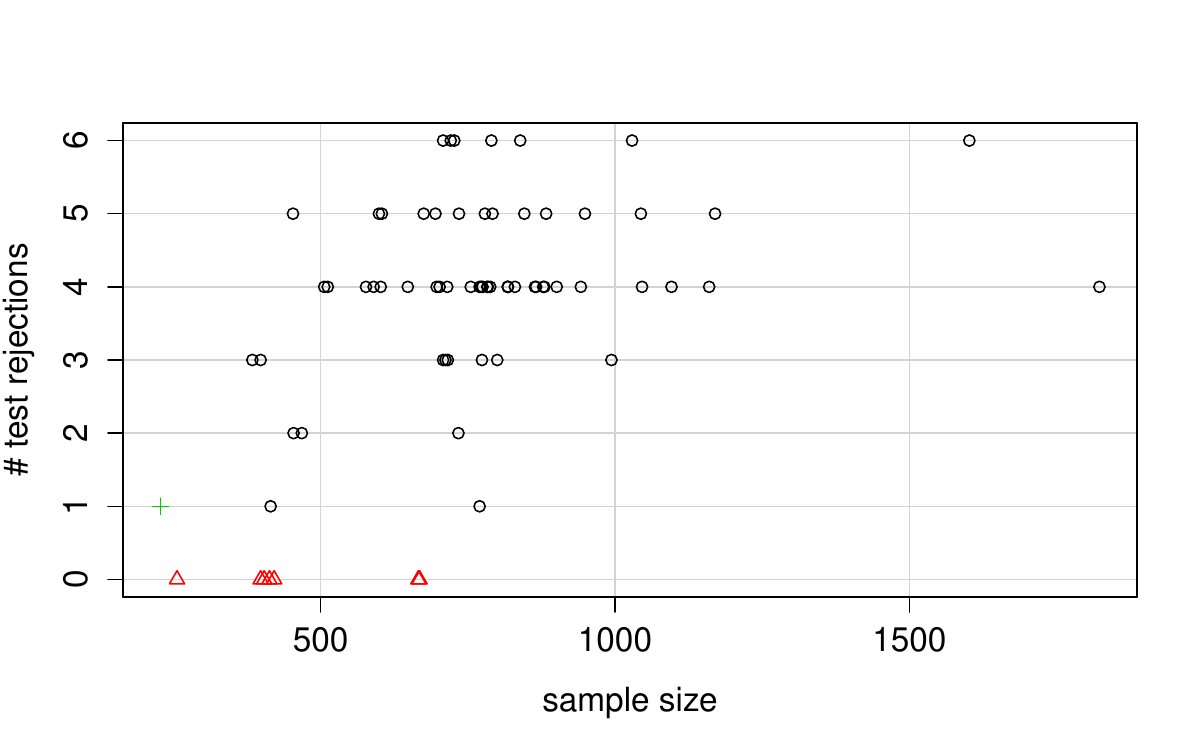}
\end{center}
\caption{Scatterplot of the number of test rejections versus the size of the subject pool. The red triangles pinpoint the 7 experiments with 0 test rejections, the green plus (+) the two overlapping laboratory experiments.\label{regSizeTestRej}}
\end{figure}

7 out of the 72 experiments exhibit zero rejections. 
From a statistical viewpoint, they all satisfy the validity condition of the test in terms of joint frequency values, that is more than 5 counts. These experiments have adjusted $p$-values for all joint frequencies significantly greater than the 5\% significance level. 
Further investigation would be required to understand the rationale of the occurrence of no test rejection in these experiments, in order to exclude spurious statistical arguments. As a first step, Figure~\ref{regSizeTestRej} plots the number of test rejections against the size of the subject pool among the experiments. 
The plot suggests that low subject pools (say, below 500) tend to reject less tests. It also suggests that it would be interesting to replicate these 7 experiments with a larger subject pool (say, 1000 or 1500), to check whether they still verify all the tests.

An important point is however the following one: all 7 polls that do satisfy all GR tests actually display no order effect (cf.~Table~\ref{tab_order-effect_7})\footnote{The data that can be obtained from Wang \textit{et al.} (2014), together with their statistical tests, easily enable to test the absence of an order effect.}. Yet, it is well-known that data sets that display no order effect can be accounted for by classical models and that quantum-like ones are not needed, whether degenerate or not. In other words, in the 7 cases in which non-degenerate quantum-like models can account for the data, quantum-like models are actually not needed. Hence, in all the 72 polls considered in this first data set, no non-degenerate quantum-like model is needed or, when needed, is possible.

\begin{table}[tb]
\caption{$p$-values for the order effect tests performed over the 7 experiments that do satisfy the GR equations. The null hypothesis of absence of order effect cannot be rejected.}
\begin{center}
\begin{tabular}{ l | c c c c c c c }
\label{tab_order-effect_7}
exp ID & 26 & 27 & 42 & 43 & 44 & 47 & 51 \\
\hline
$p$-value & 0.33 & 0.85 & 0.17 & 0.62 & 0.94 & 0.18 & 0.38 \\
\end{tabular}
\end{center}
\end{table}

\bigskip

Let us now sum up the consequences for each paper. The 6 experiments considered in Wang and Busemeyer (2013) fail at least one GR test, and thus we can safely say that they cannot be accounted for with a non-degenerate quantum-like model. For instance, their 2D figures with rays cannot be considered as representations of the model for their data, but instead should be thought as an illustrative toy model. Of course, as Wang and Busemeyer (2013) allow their model to be either degenerate or non-degenerate, there remains the possibility to account for the data with a degenerate version of the model, but our GR test is not able to pronounce on that option (cf. Section~6 for a further discussion).

Among the 66 extra experiments considered in Wang \textit{et al.} (2014), which all come from the PEW Research Center, a very large majority, 59, fail at least one test and the same consequence holds: they cannot be accounted for with a non-degenerate quantum-like model either. Similarly, as Wang \textit{et al.} (2014) allow their models to be either degenerate or non-degenerate, we do not claim here that they are wrong but only that the non-degenerate versions are not empirically adequate for these data --- there remains the theoretical possibility of degenerate versions of the model.
7 experiments remain that can be accounted for with a non-degenerate model (it would be interesting to replicate these experiments with a larger number of subjects, to check whether this still holds), but as they display no order effect, a quantum-like model is actually not needed here and a classical model could account for them.

\subsection{Second data set: Conte \textit{et al.} (2009)}
\label{subsec_res-conte}

This section considers the experiments conducted by Conte \textit{et al.} (2009) about visual perception.
The validity conditions of the statistical test are not fully met because the number of subjects per experiment is around 60, posing the issue of statistical significance for such a small sample. 
In particular, the statistical requirement of at least 5 counts for each joint frequency is not always met in this data set. However, for the sake of completeness, the statistical tests are reported here.%
\footnote{An experiment similar to the one reported in Conte \textit{et al.} (2009) is mentioned in Khrennikov (2010, p.~79-86). However, we prefer to focus on the former because in the latter (i) the number of subjects involved is still lower and (ii) as no reverse order in the questions is considered, data is lacking to fully test the GR equations.}

4 distinct experiments are performed, each comprising two treatments involving a population of 19 to 22 year-old students. The first treatment of each experiment involves a group of subjects exposed only to test $A$, whereas the second one to test $B$ and ``soon after" to test $A$. Only the second (52 subjects) and the third (64 subjects) experiments implement the same visual task (subjects have to look at ambiguous figures of animal), but exchanging tests $A$ and $B$. Therefore, by considering only the treatment/group 2 of these two experiments, the full GR equations can be tested.

Table~\ref{Res3} reports the adjusted $p$-values for the 6 tests. As the third test is rejected at the 5\% significance level, we can conclude that a quantum-like non-degenerate model does not adequately fit the experimental data.\footnote{As the GR equations are not satisfied, neither is double-stochasticity in both senses (cf. Section~\ref{subsec_link-dble-stoch}). Actually, the fact that the data of Conte \textit{et al.} (2009) fail to follow double-stochasticity is noted by Khrennikov (2010, 107 and 114).} Here, too, the small size of the sample is associated with a small number of test rejections. A more rigorous empirical test of the GR equations would require a larger subject pool.
Anyway, as the model proposed by Conte \textit{et al.} (2009) is supposed to be only non-degenerate, we can already say that it fails and cannot be said to be empirically adequate for the reported data.

\begin{table}[tb] \caption{Experiments of Conte \textit{et al.} (2009), adjusted $p$-values for the 6 tests. The rejection is highlighted at the 5\% significance level.}\label{Res3}
\begin{center}
\begin{tabular}{ l l l l l l c }
\hline
\textbf{T1}  &  \textbf{T2} &  \textbf{T3} & \textbf{T4} & \textbf{T5} & \textbf{T6} \\
\hline
4.300 & 2.404 & \textbf{0.029} & 3.782 & 0.079 & 0.268 \\ \hline
\end{tabular}
\end{center}
\end{table}

\section{Discussion: degenerate models and the future of quantum-like research on order effect}
\label{sec_discussion}

This paper has been concerned with the quantum-like models that have been recently proposed to account for question order effect. So far, many of them have been proposed in the most general way, either as non-degenerate or as degenerate, while it was suggested that degenerate models would be probably required. We settled on the project to document this empirically, by testing whether the non-degenerate versions of the models could be sufficient. To this end, we have advocated the test of novel predictions of the models, which are equivalent to the law of reciprocity: the GR equations. After performing the test against available data, the result is quite clear: non-degenerate models are not an option, being not empirically adequate or not needed. This confirms the suggestions from the literature (cf. footnote~\ref{footnote_toy-model}) that they are too simple to be empirically adequate and should be considered as toy models only.

As we have also made it clear, our results do not rule out all quantum-like models: degenerate versions of the model, to which the GR equations do not apply, have not been tested here. 
So, a natural line of research is now to turn to degenerate quantum-like models, where answers are represented by sub-spaces of dimension 2 or larger.
For instance, this is the solution adopted by Busemeyer, Wang and Lambert-Mogiliansky (2009): after showing that a 2D quantum-like model is not doubly stochastic (which implies that it does not respect our GR equations), they switch to a 4D model.
\footnote{However, note that their 4D model is not just a degenerate version of their 2D model, as it adds some time evolution of the state between the two questions. On another question than order effect, namely the account for the conjunction fallacy, Franco (2009) considers a non-degenerate 2D model only, while Busemeyer \textit{et al.} (2011) considers $n$-dimensional ones, which can be degenerate.}

A strong suggestion in favor of this line of research comes from the following 
fact. While the GR equations are violated on a vast majority of experimental data sets, the QQ equality is experimentally verified in all these cases (for the data considered in Wang and Busemeyer 2013 and Wang \textit{et al.} 2014). 
Recall that the QQ equality is valid for quantum-like models of the kind considered here, whether they are degenerate or not (cf. Section~\ref{subsec_link-QQ-equality}). It is just an empirical prediction they all have to satisfy.
Also, Wang and Busemeyer (2009, Section~6) have argued that other kinds of models, like Bayesian or Markov ones, do not generally satisfy the QQ equality, which gives good grounds to think that some quantum-like model could account for the data (cf. also Wang \textit{et al.} 2014).
As our present results exclude non-degenerate models, the hopes are now on the degenerate ones.
Overall, the following tests might be made on a data set of question order effect: 1) test the QQ equality; if it is violated, no quantum-like model (of the sort considered in this paper) can account for it; if it is not violated, it gives good grounds to think (but does not prove) that a degenerate or non-degenerate quantum-like model can account for it; 2) test the GR equations; if they are violated, no non-degenerate quantum-like model can account for the data; if they are not, one can.

However, in our view, the route of degenerate models is not necessarily an easy one, and we would like to recommend that some research be done about their empirical adequacy.
First, degenerate models are not \textit{a priori} freed from any constraint, and this is the point on which we would like to insist very much. One should not consider that, when a data set satisfies the QQ equality and violates the GR equations, the solution safely lies in a degenerate model. The moral of our paper should be taken as insisting on the need to empirically test all quantum-like models. Note that we do not just call for running more experiments; all the tests we have done here have been made \textit{on available data}, and could have been made beforehand. Importantly, empirical adequacy does not only mean applying to a vast number of empirical cases, but also, for one experimental case, making correct predictions \textit{in all respects}, and for that it is crucial to pull the threads of all possible predictions. 
Here, we have found simple and ready-to-use tests for non-degenerate models, and we suspect that some other constraints apply to degenerate models. Research is urgently needed here to find general tests. 
At least, when a degenerate model is proposed to account for some features of a data set, we suggest that all parameters be specified and that serious efforts be directed towards empirically testing this model by unveiling experimental predictions that lie outside the feature one intends to explain, 
so as to prevent later ``bad surprises".

A promising way to test degenerate models has already been indicated in the end of Section~\ref{subsec_generalizations}: degenerate eigenspaces may actually encompass non-degenerate and more fundamental eigenspaces, that correspond to fundamental elementary questions --- to be identified. 
If such elementary questions exist, and the quantum-like literature seems not to have doubts on that, all degenerate models can actually be tested against generalized GR equations for these more fundamental questions. In other words, our GR test for non-degenerate models can possibly apply at a fundamental level for all degenerate models, provided elementary questions are identified.%
\footnote{These generalized GR equations for elementary questions might help to establish the constraints that degenerate models must satisfy in general, but for non-elementary questions, that have been discussed in the above paragraph.}
We are ready to admit, however, that the fundamental dimensions and the corresponding tests can be hard to uncover.

The need to specify the elementary dimensions of the Hilbert space for degenerate models is also required by other arguments.
Introducing dimensions of degeneracy should be justified,
so as not to be accused of being just \textit{ad hoc}.
As Pothos and Busemeyer (2013, p.~215) write, ``Dimensionality is not a free parameter". 
In quantum physics for instance, degeneration 
is usually theoretically justified with a symmetry in the Hamiltonian, 
and the degeneration can be experimentally removed by introducing some factor that was not present (e.g., the spin degeneration is removed by introducing a magnetic field). For quantum-like judgment models as well, one should provide a similar theoretical justification and an experimental way of removing the degeneration. In this respect, the idea that there exists fundamental questions and rays will certainly be helpful; the work is then to find out which fundamental dimensions are hidden behind the degenerate eigenspace.

Overall, all this argues against the view that switching to degenerate models when GR equations are violated would be a natural or unquestioned move to have and that it would solve all problems. Instead, supplementary dimensions should be justified, probably with elementary questions, and other predictions of the model than those one intends to account for should be investigated and tested, possibly with generalized GR equations.

Beyond degenerate quantum-like models in the flavor of the model presented in Section~2, there are also other possible lines of research around quantum ideas. 
For instance, Khrennikov has proposed to consider hyperbolic Hilbert space models, instead of complex ones like in this paper. Hyperbolic models enable probabilities that are not constrained in the same way, and can account for some data that are out of reach for complex quantum-like models (see Khrennikov 2010 for a synthesis).
Also, other tools than projectors, like POVMs, could be considered; 
or other axioms of standard quantum-like models, like the projection postulate or the Born rule, could be abandoned or modified.

To conclude, quantum-like models have brought to discussion many provoking and seminal ideas, such as the hypotheses that judgments or preferences might be undetermined instead of only unknown, or that non-classical probabilities could be considered. 
We have insisted on the need to empirically test existing models for question order effect, and our results suggest that non-degenerate quantum-like models should be considered more as toy models than as empirically adequate models, in accordance with suggestions from the literature, and that future investigations should focus on degenerate models.
In spite of the potential difficulties that such models face, there is plenty of room for future research in the area, and besides our results can be taken as posing a challenge liable to trigger a revival of the discussion on quantum-like models of various sorts.

\section*{Acknowledgments} 

Many thanks to Camille Aron,  Jerome R. Busemeyer, Sophie Dubois, Dorian Jullien, Corrado Lagazio, Ariane Lambert-Mogiliansky, Gabriel Lemarié, Zheng J. Wang, and to the audience from conferences at the MSHS du Sud-Est (Nice, France) and at the Toronto Fields Institute (Canada) for valuable comments or suggestions. Of course, we are solely responsible for our analysis and conclusions. We further thank Zheng J. Wang, Tyler Solloway, Richard M. Shiffrin, and Jerome R. Busemeyer who kindly gave us the opportunity to use their data set to perform our tests.

\section*{References}

\newlength{\oldparindent}
\setlength{\oldparindent}{\parindent}

\setlength{\parindent}{-.8cm}
\begin{changemargin}{.8cm}{0cm}


\textsc{Aerts}, Diederik, Jan \textsc{Broekaert}, Marek \textsc{Czachor} and Bart \textsc{D'Hooghe} (2011), ``A Quantum-Conceptual Explanation of Violations of Expected Utility in Economics", \textit{Quantum Interaction. Lecture Notes in Computer Science} 7052: 192-198.

\textsc{Aerts}, Diederik, Liane \textsc{Gabora} and Sandro \textsc{Sozzo} (2013), ``Concepts and Their Dynamics: A Quantum-Theoretic Modeling of Human Thought", \textit{Topics in Cognitive Sciences} 5: 737--772.

\textsc{Allais}, Maurice (1953), ``Le comportement de l'homme rationnel devant le risque : critique des postulats et axiomes de l'école Américaine", \textit{Econometrica} 21:503--546.

\textsc{Ashby}, F. Gregory and Nancy A. \textsc{Perrin} (1988), ``Towards a unified theory of similarity and recognition", \textit{Psychological Review}, 95:124--50.

\textsc{Ashtiani}, Mehrdad and Mohammad A. \textsc{Azgomi} (2015), ``A survey of quantum-like approaches to decision making and cognition", \textit{Mathematical Social Sciences} 75: 49--80.

\textsc{Atmanspacher}, Harald and Hartmann \textsc{Römer} (2012), ``Order effects in sequential measurements of non-commuting psychological observables", \textit{Journal of Mathematical Psychology} 56: 274-280.

\textsc{Busemeyer}, Jerome R. and Peter D. \textsc{Bruza} (2012), \textit{Quantum models of Cognition and Decision}, Cambridge: Cambridge University Press.

\textsc{Busemeyer}, Jerome R., Emmanuel M. \textsc{Pothos}, Riccardo \textsc{Franco} and Jennifer S. \textsc{Trueblood} (2011), ``A Quantum Theoretical Explanation for Probability Judgment Errors", \textit{Psychological Review} 118 (2): 193--218.

\textsc{Busemeyer}, Jerome R., Zheng J. \textsc{Wang} (2007), ``Quantum information processing explanation for interactions between inferences and decisions", in: D. Bruza, W. Lawless, K. van Rijsbergen and D.A. Sofge (eds.), \textit{Quantum Interaction: Papers from the AAAI Spring Symposium}, Technical Report SS-07-08, AAAI Press, p. 91–97.

\textsc{Busemeyer}, Jerome R., Zheng J. \textsc{Wang}, and Ariane \textsc{Lambert-Mogiliansky} (2009), ``Empirical comparison of Markov and quantum models of decision making", \textit{Journal of Mathematical Psychology} 53: 423--433.

\textsc{Busemeyer}, Jerome R., Zheng J. \textsc{Wang}, Emmanuel M. \textsc{Pothos} and Jennifer S. \textsc{Trueblood} (2015), ``The Conjunction Fallacy, Confirmation, and Quantum Theory: Comment on Tentori, Crupi, and Russo (2013)", \textit{Journal of Experimental Psychology: General} 144(1): 236-243.

\textsc{Busemeyer}, Jerome R., Zheng J. \textsc{Wang}, and James T. \textsc{Townsend} (2006), ``Quantum dynamics of human decision-making", \textit{Journal of Mathematical Psychology} 50 (3): 220-241.

\textsc{Conte}, Elio, Andrei Yuri \textsc{Khrennikov}, Orlando \textsc{Todarello}, Antonio \textsc{Federici}, Leonardo \textsc{Mendolicchio}, and Joseph P. \textsc{Zbilut} (2009), ``Mental States Follow Quantum Mechanics During Perception and Cognition of Ambiguous Figures", \textit{Open Systems \& Information Dynamics} 16(1) 1--17.

\textsc{Danilov}, Vladimir I. and Ariane \textsc{Lambert-Mogiliansky} (2008), ``Measurable systems and behavioral sciences", \textit{Mathematical Social Sciences} 55(3): 315--340.

\textsc{Danilov}, Vladimir I. and Ariane \textsc{Lambert-Mogiliansky} (2010), ``Expected utility theory under non-classical uncertainty", \textit{Theory and Decision} 68(1): 25--47.

\textsc{Ellsberg}, Daniel (1961), ``Risk, ambiguity, and the Savage axioms", \textit{Quarterly Journal of Economics} 75(4): 643--669.

\textsc{Franco}, Riccardo (2009), ``The conjunction fallacy and interference effects", \textit{Journal of Mathematical Psychology} 53: 415--422.

\textsc{Gavanski}, Igor and David R. \textsc{Roskos-Ewoldsen} (1991), ``Representativeness and conjoint
probability", \textit{Journal of Personality and Social Psychology} 61:181--94.


\textsc{Khrennikov}, Andrei (2010), \textit{Ubiquitous Quantum Structure. From Psychology to Finance}, Heildelberg: Springer.

\textsc{Khrennikov}, Andrei and Emmanuel \textit{Haven} (2009), ``Quantum mechanics and violations of the sure-thing principle: The use of probability interference and other concepts", \textit{Journal of Mathematical Psychology} 53: 378-388.

\textsc{Krumhansl}, Carol L. (1978), ``Concerning the applicability of geometric models to similarity data: The interrelationship between similarity and spatial density", \textit{Psychological Review} 85:445--63.

\textsc{Lambert-Mogiliansky}, Ariane,  Shmuel \textsc{Zamir} and Hervé \textsc{Zwirn} (2009), ``Type indeterminacy: A model of the KT(Kahneman–Tversky)-man", \textit{Journal of Mathematical Psychology} 53:349--361.

\textsc{Moore}, David W. (2002), ``Measuring new types of question-order effects", \textit{Public Opinion Quarterly}, 66: 80--91.

\textsc{Peres}, Asher (1993), \textit{Quantum Theory: Concepts and Methods}, Kluwer.

\textsc{Pothos}, Emmanuel M.  and Jerome R. \textsc{Busemeyer} (2009), ``A quantum probability explanation for violations of `rational' decision theory", \textit{Proceedings of the Royal Society B} 276:2171-2178.

\textsc{Pothos}, Emmanuel M.  and Jerome R. \textsc{Busemeyer} (2011), ``A quantum probability explanation for violations of symmetry in similarity judgments", \textit{Proceedings of the 32nd Annual conference of the Cognitive Science Society} 2848-2854.

\textsc{Pothos}, Emmanuel M.  and Jerome R. \textsc{Busemeyer} (2013), ``Can quantum probability provide a new direction for cognitive modeling?", \textit{Behavioral and Brain Sciences} 36: 255--327.

\textsc{Schuman}, Howard and Stanley \textsc{Presser} (1981), \textit{Questions and answers in attitude surveys: Experiments on question form, wording, and context}, Academic Press.

\textsc{Shafir}, Eldar and Amos \textsc{Tversky} (1992), ``Thinking through uncertainty: nonconsequential reasoning and choice", \textit{Cognitive Psychology} 24:449--74.

\textsc{Stolarz-Fantino}, Stepanie, Edmund \textsc{Fantino}, Daniel J. \textsc{Zizzo}, and Julie \textsc{Wen} (2003), ``The conjunction effect: New evidence for robustness", \textit{American Journal of Psychology} 116(1):15--34.

\textsc{Tourangeau}, Roger, Lance J. \textsc{Rips} and Kenneth A. \textsc{Rasinski} (2000),  \textit{The psychology of survey response}, Cambridge University Press.

\textsc{Trueblood}  Jennifer S. and Jerome R. \textsc{Busemeyer} (2011), ``A Quantum Probability Account of Order Effects in Inference", \textit{Cognitive Science} 35: 1518--1552.

\textsc{Tversky}, Amos (1977), ``Features of similarity", \textit{Psychological Review} 84(4): 327--52.

\textsc{Tversky}, Amos and Daniel \textsc{Kahneman} (1982), ``Judgments of and by representativeness", in: \textit{Judgment under uncertainty: Heuristics and biases}, ed. by D. Kahneman, P. Slovic
\& A. Tversky, p. 84–98, Cambridge University Press.

--- (1983), ``Extensional versus intuitive reasoning: The conjunction fallacy in probability judgment", \textit{Psychological Review} 90(4): 293--315.

\textsc{Wang}, Zheng J. and Jerome R. \textsc{Busemeyer} (2013), ``A quantum question order model supported by empirical tests of an a priori and precise prediction", \textit{Topics in Cognitive Science} 5(4): 689--710.

\textsc{Wang}, Zheng J., Tyler \textsc{Solloway}, Richard M. \textsc{Shiffrin} and Jerome R. \textsc{Busemeyer} (2014),  ``Context effects produced by question orders reveal quantum nature of human judgments", \textit{PNAS}, \url{www.pnas.org/cgi/doi/10.1073/pnas.1407756111}.

\textsc{Wilson}, David C., David W. \textsc{Moore}, Patrick F. \textsc{McKay}, and Derek R. \textsc{Avery} (2008), ``Affirmative action programs for women and minorities", \textit{Public Opinion Quarterly} 72(3), 514-522.

\textsc{Yukalov}, Vyacheslav I. and Didier \textsc{Sornette} (2011), ``Decision theory with prospect interference and entanglement", \textit{Theory and Decision}, 70: 283--328.

\end{changemargin}

\setlength{\parindent}{\oldparindent}

\newpage

\appendix
\section{Testing the equality of two conditional relative frequencies}\label{AppA}
The statistical test is to compare two conditional relative frequencies $y$ and $x$, with the null hypothesis that they are equal. 
The test is therefore
\begin{equation}
y=x \label{yxTest},
\end{equation}
where both $y$ and $x$ are observed conditional relative frequencies.\\
Testing equation \ref{yxTest} is equivalent to test
\begin{equation*}
\log \left(\frac{y}{1-y} \right)=\log \left(\frac{x}{1-x} \right),
\end{equation*}
given that $y$ and $x$ are not equal to zero.\\
Alternatively, we can formulate the test in terms of the log odds ratio (OR)
\begin{equation*}
\log({\rm OR}) = \log \left(\frac{\frac{y}{1-y}}{\frac{x}{1-x}} \right) = 0.
\end{equation*}
Now let us suppose that we want to test\\ 
\begin{equation}\label{condProbTest} 
f(a_0|b_0) = f(b_0|a_0). 
\end{equation}
We can thus test the following condition:
\begin{equation*}
\log\left(\frac{f(a_0| b_0)}{1- f(a_0|b_0)}\right) = \log\left(\frac{f(b_0| a_0)}{1-f(b_0|a_0)}\right), 
\end{equation*}
or
\begin{equation*}
\log\left(\frac{f(a_0| b_0)}{f(a_1|b_0)}\right) = \log\left(\frac{f(b_0| a_0)}{f(b_1|a_0)}\right).
\end{equation*}
By expressing the conditional relative frequencies in terms of joint frequencies, that is,
\begin{equation*}  
f(a_0|b_0) = \frac{n(b_0, a_0)}{n(b_0, \cdot)}, \ f(b_0|a_0) = \frac{n(a_0, b_0)}{n(a_0, \cdot)}, \dots
\end{equation*}
being ${n(b_0, \cdot)}$ and ${n(a_0, \cdot)}$ the marginal frequencies of $\mathbf{B}$ and $\mathbf{A}$, respectively, we obtain
\begin{equation*}
\log\left(\frac{n(b_0, a_0)}{n(b_0, \cdot)}\frac{n(b_0, \cdot)}{n(b_0, a_1)}\right) = \log\left(\frac{n(a_0, b_0)}{n(a_0, \cdot )}\frac{n(a_0, \cdot)}{n(a_0, b_1)}\right),
\end{equation*}
or simplifying
\begin{equation} \label{logORTest}
\log\left(\frac{n(b_0, a_0)n(a_0, b_1)}{n(b_0, a_1)n(a_0, b_0)}\right)=0.
\end{equation}
We can thus test indifferently eq.~\ref{condProbTest} or \ref{logORTest}.

\newpage

\section{Results of the tests for the first data set}
\label{sec_appendix_results}

\begin{longtable}{ l l l l l l c c}
\caption{Adjusted $p$-values of the 6 tests for the 72 experiments of the first data set. 
The rejection is highlighted at the 5\% significance level. The 7$^{th}$ column reports the number of rejections (\#R) at the 5\% significance level, and the 8$^{th}$ one reports the experiment ID (cf. Table~\ref{tab_questions-data-set}).}\label{longtableres}\\
\hline
\textbf{T1}  & \textbf{T2}&  \textbf{T3} & \textbf{T4} & \textbf{T5} & \textbf{T6} & \textbf{\#R at 5\%} & \textbf{exp ID}\\
\hline
\endfirsthead
\multicolumn{8}{l}%
{\tablename\ \thetable\ -- \textit{Continued from previous page}} \\
\hline
\textbf{T1}  & \textbf{ T2 }&  \textbf{T3} & \textbf{T4} & \textbf{T5} & \textbf{T6} & \textbf{\#R at 5\%} & \textbf{exp. ID}\\
\hline
\endhead
\hline \multicolumn{8}{l}{\textit{Continued on next page}} \\
\endfoot
\hline
\endlastfoot
\textbf{0.011} & \textbf{0.000} &         4.718  &         0.057  & \textbf{0.023} & \textbf{0.000} & 4 & 1  \\
\textbf{0.000} & \textbf{0.004} &         2.652  &         5.163  & \textbf{0.000} & \textbf{0.001} & 4 & 2  \\
        0.135  &         0.975  &         0.321  &         3.828  & \textbf{0.000} & \textbf{0.033} & 2 & 3  \\
\textbf{0.000} & \textbf{0.000} &         0.113  &         0.191  &         0.078  & \textbf{0.000} & 3 & 4  \\
\textbf{0.000} & \textbf{0.000} &         1.653  &         1.151  & \textbf{0.008} & \textbf{0.000} & 4 & 5  \\
\textbf{0.000} & \textbf{0.000} &         4.577  &         2.138  & \textbf{0.000} & \textbf{0.000} & 4 & 6  \\
\textbf{0.000} & \textbf{0.000} &         1.400  & \textbf{0.023} & \textbf{0.006} & \textbf{0.000} & 5 & 7  \\
\textbf{0.000} & \textbf{0.000} & \textbf{0.001} &         1.279  &         0.180  & \textbf{0.001} & 4 & 8  \\
\textbf{0.021} & \textbf{0.000} &         0.181  & \textbf{0.024} &         1.682  & \textbf{0.000} & 4 & 9  \\
\textbf{0.001} & \textbf{0.000} &         0.274  & \textbf{0.006} &         0.096  & \textbf{0.000} & 4 & 10 \\
        0.749  & \textbf{0.000} &         0.066  & \textbf{0.000} &         2.517  & \textbf{0.000} & 3 & 11 \\
\textbf{0.000} & \textbf{0.000} &         3.319  &         0.607  & \textbf{0.000} & \textbf{0.000} & 4 & 12 \\
\textbf{0.000} & \textbf{0.000} &         0.102  &         1.728  & \textbf{0.000} & \textbf{0.000} & 4 & 13 \\
\textbf{0.000} & \textbf{0.000} &         1.506  &         5.892  & \textbf{0.000} & \textbf{0.000} & 4 & 14 \\
\textbf{0.000} & \textbf{0.000} & \textbf{0.002} &         1.050  & \textbf{0.000} & \textbf{0.000} & 5 & 15 \\
\textbf{0.000} & \textbf{0.000} & \textbf{0.032} & \textbf{0.033} & \textbf{0.000} & \textbf{0.000} & 6 & 16 \\
\textbf{0.000} & \textbf{0.000} & \textbf{0.000} & \textbf{0.014} & \textbf{0.000} & \textbf{0.000} & 6 & 17 \\
\textbf{0.000} & \textbf{0.000} & \textbf{0.000} & \textbf{0.001} & \textbf{0.002} & \textbf{0.000} & 6 & 18 \\
\textbf{0.000} & \textbf{0.000} & \textbf{0.004} & \textbf{0.019} & \textbf{0.000} & \textbf{0.000} & 6 & 19 \\
\textbf{0.000} & \textbf{0.000} &         0.535  &         0.082  & \textbf{0.001} & \textbf{0.000} & 4 & 20 \\
\textbf{0.000} & \textbf{0.000} & \textbf{0.000} & \textbf{0.000} & \textbf{0.000} & \textbf{0.000} & 6 & 21 \\
\textbf{0.000} & \textbf{0.000} & \textbf{0.001} & \textbf{0.000} &         0.354  & \textbf{0.000} & 5 & 22 \\
\textbf{0.000} & \textbf{0.000} & \textbf{0.000} & \textbf{0.000} & \textbf{0.025} & \textbf{0.000} & 6 & 23 \\
\textbf{0.000} & \textbf{0.000} &         0.107  & \textbf{0.037} & \textbf{0.000} & \textbf{0.000} & 5 & 24 \\
\textbf{0.000} & \textbf{0.000} & \textbf{0.000} & \textbf{0.000} & \textbf{0.002} & \textbf{0.000} & 6 & 25 \\
        4.172  &         4.213  &         1.406  &         2.708  &         2.720  &         0.728  & 0 & 26 \\
        5.811  &         4.289  &         1.423  &         4.154  &         1.408  &         2.316  & 0 & 27 \\
        4.261  &         0.150  & \textbf{0.020} &         0.450  &         0.096  &         3.540  & 1 & 28 \\
        1.557  & \textbf{0.039} &         0.072  & \textbf{0.001} & \textbf{0.003} &         4.611  & 3 & 29 \\
        5.368  & \textbf{0.026} &         0.070  & \textbf{0.001} & \textbf{0.001} &         2.092  & 3 & 30 \\
        0.135  &         1.870  &         0.427  & \textbf{0.005} & \textbf{0.000} &         2.286  & 2 & 31 \\
        0.912  & \textbf{0.000} & \textbf{0.000} & \textbf{0.000} & \textbf{0.000} &         3.896  & 4 & 32 \\
\textbf{0.011} &         2.285  & \textbf{0.002} &         0.068  & \textbf{0.000} & \textbf{0.000} & 4 & 33 \\
\textbf{0.012} &         1.420  & \textbf{0.002} & \textbf{0.000} & \textbf{0.000} & \textbf{0.021} & 5 & 34 \\
        0.580  &         0.074  & \textbf{0.000} & \textbf{0.000} & \textbf{0.000} & \textbf{0.038} & 4 & 35 \\
        1.093  &         1.382  & \textbf{0.041} & \textbf{0.035} & \textbf{0.000} &         0.427  & 3 & 36 \\
        2.913  &         0.073  & \textbf{0.000} & \textbf{0.001} & \textbf{0.000} &         0.107  & 3 & 37 \\
        0.179  & \textbf{0.000} & \textbf{0.000} & \textbf{0.000} & \textbf{0.000} &         1.062  & 4 & 38 \\
        1.373  & \textbf{0.000} & \textbf{0.000} & \textbf{0.000} & \textbf{0.000} &         2.482  & 4 & 39 \\
        4.564  & \textbf{0.000} & \textbf{0.000} & \textbf{0.000} & \textbf{0.000} &         4.084  & 4 & 40 \\
        1.936  & \textbf{0.000} & \textbf{0.000} & \textbf{0.000} & \textbf{0.000} &         5.716  & 4 & 41 \\
        5.188  &         1.343  &         1.723  &         1.649  &         2.101  &         5.229  & 0 & 42 \\
        2.651  &         5.452  &         5.452  &         2.212  &         2.212  &         6.000  & 0 & 43 \\
        3.965  &         4.249  &         3.554  &         2.414  &         1.999  &         5.064  & 0 & 44 \\
\textbf{0.001} & \textbf{0.000} & \textbf{0.000} & \textbf{0.000} &         0.546  &         0.055  & 4 & 45 \\
\textbf{0.000} & \textbf{0.000} &         0.860  &         4.837  & \textbf{0.000} & \textbf{0.000} & 4 & 46 \\
        0.050  &         0.181  &         3.527  &         4.402  &         0.523  &         1.039  & 0 & 47 \\
\textbf{0.004} & \textbf{0.044} &         2.358  &         4.837  & \textbf{0.036} &         0.212  & 3 & 48 \\
\textbf{0.005} &         0.347  &         2.185  &         1.999  &         0.103  &         1.537  & 1 & 49 \\
\textbf{0.000} &         4.281  & \textbf{0.000} & \textbf{0.000} & \textbf{0.000} & \textbf{0.000} & 5 & 50 \\
        1.939  &         1.910  &         3.856  &         5.503  &         3.163  &         3.307  & 0 & 51 \\
        0.141  & \textbf{0.000} & \textbf{0.000} & \textbf{0.000} & \textbf{0.025} &         0.377  & 4 & 52 \\
\textbf{0.003} &         4.472  & \textbf{0.003} & \textbf{0.003} & \textbf{0.000} & \textbf{0.023} & 5 & 53 \\
\textbf{0.035} & \textbf{0.000} & \textbf{0.000} & \textbf{0.000} & \textbf{0.000} &         0.060  & 5 & 54 \\
        5.498  & \textbf{0.000} & \textbf{0.000} & \textbf{0.000} & \textbf{0.000} &         4.845  & 4 & 55 \\
\textbf{0.013} &         0.125  &         3.852  &         2.282  & \textbf{0.046} &         0.360  & 2 & 56 \\
\textbf{0.000} & \textbf{0.000} &         1.843  &         4.369  & \textbf{0.000} & \textbf{0.000} & 4 & 57 \\
        1.922  & \textbf{0.000} & \textbf{0.000} & \textbf{0.000} & \textbf{0.000} & \textbf{0.043} & 5 & 58 \\
\textbf{0.001} &         3.012  &         0.389  & \textbf{0.026} & \textbf{0.000} &         0.179  & 3 & 59 \\
\textbf{0.000} &         1.570  & \textbf{0.001} & \textbf{0.000} & \textbf{0.000} & \textbf{0.000} & 5 & 60 \\
\textbf{0.000} & \textbf{0.000} &         0.128  &         4.872  & \textbf{0.000} & \textbf{0.000} & 4 & 61 \\
\textbf{0.003} & \textbf{0.000} & \textbf{0.001} & \textbf{0.004} &         0.608  &         2.054  & 4 & 62 \\
\textbf{0.000} & \textbf{0.000} &         5.106  &         3.456  & \textbf{0.000} & \textbf{0.000} & 4 & 63 \\
\textbf{0.000} & \textbf{0.000} & \textbf{0.000} &         3.333  & \textbf{0.012} & \textbf{0.000} & 5 & 64 \\
\textbf{0.021} &         4.789  &         0.065  & \textbf{0.001} & \textbf{0.000} & \textbf{0.039} & 4 & 65 \\
\textbf{0.002} &         0.466  & \textbf{0.027} &         0.135  & \textbf{0.000} & \textbf{0.000} & 4 & 66 \\
\textbf{0.000} & \textbf{0.000} &         2.771  & \textbf{0.024} & \textbf{0.011} & \textbf{0.000} & 5 & 67 \\
\textbf{0.000} & \textbf{0.000} &         2.909  &         0.075  & \textbf{0.000} & \textbf{0.000} & 4 & 68 \\
\textbf{0.020} & \textbf{0.000} & \textbf{0.000} & \textbf{0.001} & \textbf{0.004} &         3.581  & 5 & 69 \\
        4.481  &         0.130  &         0.709  & \textbf{0.038} &         0.293  &         2.110  & 1 & 70 \\
        0.511  & \textbf{0.008} &         4.589  &         0.332  &         1.616  &         0.115  & 1 & 71 \\
\textbf{0.016} & \textbf{0.000} & \textbf{0.010} & \textbf{0.014} &         0.654  &         0.479  & 4 & 72
\end{longtable}

\end{document}